\documentclass[journal]{IEEEtran}
\ifCLASSINFOpdf
\else
\usepackage[dvips]{graphicx}
\fi
\usepackage[cmex10]{amsmath}
\allowdisplaybreaks[4]
\newtheorem{theorem}{Theorem}
\usepackage{upgreek}
\usepackage{booktabs}
\usepackage{geometry}
\usepackage{epsfig}
\usepackage{latexsym}
\usepackage{soul}
\usepackage{multirow}
\usepackage{stfloats}
\usepackage{epstopdf}
\usepackage{color}  
\usepackage{makecell}
\usepackage{tabularx} 
\usepackage{amssymb}
\usepackage{enumerate}
\graphicspath{{./Figures/}}
\usepackage{color}
\usepackage{bbm}
\usepackage{bm}
\usepackage{cite}
\usepackage[tight,footnotesize]{subfigure}
\usepackage{balance}
\usepackage{mathrsfs}
\usepackage{verbatim}
\usepackage{dsfont}
\usepackage{verbatim}
\usepackage{algorithm}
\usepackage{algorithmic}
\usepackage{tikz}
\usepackage{diagbox}
\usepackage{caption}
\usepackage[framemethod=tikz]{mdframed}
\usepackage{multicol}
\usepackage{environ}
\usepackage{tikz}
\usepackage{subcaption}
\begin{document}
\title{Hybrid Beamforming with Widely-spaced-array for Multi-user Cross-Near-and-Far-Field Communications}
\author{Heyin~Shen, Yuhang Chen, Chong~Han,~\IEEEmembership{Senior Member,~IEEE,}
and~Jinhong~Yuan,~\IEEEmembership{Fellow,~IEEE}
\thanks{Heyin Shen, Yuhang Chen and Chong Han are with the Terahertz Wireless Communications (TWC) Laboratory, Shanghai Jiao Tong University, Shanghai 200240, China (e-mail: heyin.shen@sjtu.edu.cn; yuhang.chen@sjtu.edu.cn; chong.han@sjtu.edu.cn).}
\thanks{Jinhong Yuan is with the School of Electrical Engineering and Telecommunications, University of New South Wales, Sydney, NSW 2052, Australia (e-mail: j.yuan@unsw.edu.au).}
}

\markboth{}
\MakeLowercase
\maketitle
\begin{abstract}
\boldmath With multi-GHz bandwidth, Terahertz (THz) beamforming has drawn increasing attention in the sixth generation (6G) and beyond communications. Existing beamforming designs mainly focus on a compact antenna array where typical communication occurs in the far-field. However, in dense multi-user scenarios, only relying on far-field angle domain fails to distinguish users at similar angles. Therefore, a multi-user widely-spaced array (MU-WSA) is exploited in this paper, which enlarges the near-field region to introduce the additional distance domain, leading to a new paradigm of cross-near-and-far-field (CNFF) communication. Under this paradigm, the CNFF channel model is investigated, based on which the subarray spacing $d_s$ and the number of subarrays $K$ in MU-WSA are optimized to maximize the channel capacity. Then, in sub-connected (SC) systems, an subarray-based alternating optimization (S-AO) beamforming algorithm is proposed to deal with the special block-diagonal format of the analog precoder. For fully-connected (FC) systems, a low-complexity steering vector reconstruction (SVR)-based algorithm is proposed by constructing specialized steering vectors of MU-WSA. Numerical evaluations show that due to distance domain resolutions, the MU-WSA can improve the SE by over $60$\% at a power of $20$~dBm compared to the compact array. Additionally, the proposed S-AO algorithm in the SC system can achieve over 80\% of the sum (SE) of the FC system while reducing the number of phase shifters by $K^2$, thereby lowering power consumption. The SVR algorithm in the FC system can achieve over 95\% of the upper bound of SE, but it takes only 10\% of the running time of the singular value decomposition (SVD)-based algorithms.
\end{abstract}

\IEEEpeerreviewmaketitle
\section{Introduction}
\label{section_intro}
Terahertz (THz) communication has been regarded as an enabling technology for 6G and beyond wireless networks thanks to its ultra-broad continuous bandwidth, ranging from $0.1$ to $10$ THz~\cite{overview}. However, due to much higher carrier frequency, the severe path loss at the THz band significantly limits the transmission distance and the coverage range~\cite{combat}. Fortunately, by exploiting the sub-millimeter wavelength, the ultra-massive multiple-input multiple-output (UM-MIMO) antenna array can be utilized as a potential solution to achieve high array gain and compensate for the propagation loss~\cite{UM-MIMO}. However, with traditional fully-digital (FD) beamforming architectures where each antenna is equipped with one dedicated RF chain, the huge antenna array brings high hardware complexity and power consumption~\cite{Challenges}. To tackle this problem, hybrid beamforming with fewer RF chains is proposed as an appealing technology, which is able to achieve near-optimal performance with acceptable hardware complexity~\cite{Survey}. 

Most existing hybrid beamforming schemes consider far-field transmission where the electromagnetic (EM) wavefront is approximated as a plane such that the planar wave model (PWM) is utilized to characterize the channel~\cite{modeling}. Under this assumption, the fully-connected (FC) and the sub-connected (SC) architectures are widely-explored. In FC where each RF chain connects to all the antennas, near-optimal spectral efficiency (SE) can be obtained~\cite{Spacespartial,HAD,HBD,HBF}. While in SC where each RF chain only connects to a subset of antennas, the power consumption and the hardware complexity can be reduced at the cost of SE~\cite{subconnected,Dual,energy_sub,subconnected1,subconnected2}.

However, only considering the far-field PWM imposes two limitations for THz multi-user communications. On one hand, the number of transmit data streams for each user is upper-bounded by the number of multipath components, which could be very limited at the THz band, e.g., less than 5~\cite{InterIntra,Multi-Ray,measurement}. On the other hand, as far-field beam steering can only provide angle domain resolution, it cannot distinguish users located at the same or similar angles~\cite{multifocus,different}. Consequently, far-field beamforming lacks the ability to support ubiquitous connectivity in the 6G vision, e.g., more than $10^6 /\text{km}^2$ \cite{opportunistic}. To tackle these problems, near-field beamforming has been proposed to provide additional spatial degrees of freedom (SDoFs) brought by the non-negligible spherical wavefront in near-field~\cite{beam_focusing,Multiple-access}. Leveraging this property, the authors in~\cite{beam_focusing} proposed several multi-users beam focusing schemes for fully digital, hybrid, and DMA architectures, respectively. Moreover, the authors in~\cite{Multiple-access} proposed the location division multiple access (LDMA) where focused beams can be generated to point towards a specific location rather than just a particular direction as the far-field beam steering does, thus increasing the spatial multiplexing gain, and enabling distance level resolution for multi-user communications.

Despite the advantages of near-field transmission, conventional compact antenna arrays usually provide a small near-field region due to the extremely small wavelength at the THz band. Consider a uniform planar array (UPA) with $N$ antennas, the array aperture can be calculated as $S = (\sqrt{N}-1)\sqrt{2}\lambda /2$, where $\lambda$ denotes the wavelength. As the classical boundary between the far-field and the near-field, the Rayleigh distance can then be calculated as $D_{ray}=\frac{2S^2}{\lambda}=(\sqrt{N}-1)^2\lambda \approx N\lambda$~\cite{unified}. Namely, the Rayleigh distance is approximately proportional to the product of the number of antennas and the wavelength. For example, UPA architecture with 1024 antennas at 0.3 THz only provides a near-field region around $D_{ray}\approx 1024 \times 0.001 = 1$~m. Therefore, wireless communication usually operates in the far-field under this practical consideration. 

Several studies of near-field beamforming assume extremely large number of antenna elements~\cite{RIS-4096,near_field_MU,RIS-2873} or with uniform linear array (ULA)~\cite{wideband-nearfield,Parallel,rainbow} to increase the near-field region. However, incorporating too many antennas within a limited array space leads to complex hardware configurations, while utilizing ULA may restrict the available spatial dimensions. Another possible solution to endure a large number of antennas and to increase the spatial multiplexing is the distributed antenna system (DAS)~\cite{DAS1,DAS2}. However, DAS usually requires complicated user association strategy and careful design of coordination between distributed antenna arrays, which can increase the complexity of signal processing~\cite{DAS3}.

Instead, a promising beamforming architecture for this problem is to utilize the widely-spaced array (WSA) architecture, which is also known as sparse array, or line-of-sight (LoS) MIMO~\cite{LoS_MIMO,LoS_MIMO2,sparsearray1}. In our previous work, WSA has been proposed for point-to-point (P2P) communication~\cite{InterIntra}. Specifically, the antennas at both the transmitter and the receiver are uniformly divided into $K$ widely-spaced subarrays, i.e., the subarray spacing is larger than half-wavelength. As a result, the effective array aperture can be increased without introducing numerous antenna elements, thus enlarging the near-field region while maintaining an acceptable hardware complexity. Moreover, compared to DAS, WSA with one common site can easily support joint beamforming without complex inter-site coordination~\cite{modular}. Based on WSA, the authors in~\cite{modular} derived a closed-form expression of the maximum SNR at the BS in uplink systems with one single-antenna receiver. When it comes to multi-user scenarios, the available size of user equipment and the number of receiving antennas are often significantly constrained, making it impractical to adopt the WSA. In this case, the authors in~\cite{MU-modular} focused on an uplink system with multiple single-antenna users, where the beam focusing pattern was analyzed and a user grouping strategy was designed. However, in downlink multi-user systems, a new diagram of cross-field communications occurs where users residing in both near- and far-field are served simultaneously. In this context, the design of the antenna array, including the number of subarrays and the subarray spacing, as well as the beamforming strategy under different connection types, i.e., fully-connected and sub-connected structures, remain unexplored in the literature.

To fill this research gap, in this paper we design the hybrid beamforming scheme for downlink multi-user WSA (MU-WSA), where we analyze both the FC and the SC cases. The distinctive contributions of this work are summarized as follows.
\begin{itemize}
  \item We analyze the WSA system in multi-user scenarios where the base station (BS) is equipped with widely-spaced subarrays such that the near-field region is enlarged without increasing the number of transmit antennas. While at each user, the compact array is adopted to fit the limited size of the user equipment. In this system, a cross-near-and-far-field (CNFF) channel model is established as a simplified yet accurate approximation of the spherical-wave model (SWM). Then, we analyze the influence of the number of subarrays and the subarray spacing on the sum SE. Specifically, we prove that for a fixed number of subarrays, the channel capacity increases with a larger subarray spacing. Based on this result, we derive the closed-form expression of the optimal subarray spacing.
  \item We investigate the sub-connected MU-WSA with multi-antenna users. Specifically, we first divide the design problem into the analog and digital stages where the block-diagonalization (BD) method is applied in the digital stage to mitigate the user interference. For the analog precoder, we consider a more general case such that each subarray in MU-WSA can be connected to multiple RF chains rather than only one RF chain as in most SC studies~\cite{subconnected,subconnected1,subconnected2,energy_sub,Dual}. In this context, we formulate the analog precoder as a special block diagonal matrix where each block is itself a matrix. Then, based on mathematical derivations, we decompose the contribution of each submatrix, and propose a subarray-based alternating optimization (S-AO) algorithm to iteratively optimize the precoder for each subarray.
  
  \item We investigate the fully-connected MU-WSA where we propose a novel low-complexity subarray-based steering vector reconstruction (SVR) method, which extends the conventional far-field steering method to the MU-WSA architecture. By leveraging the CNFF channel model, the new beam steering vector for all subarrays is derived in the MU-WSA, using only the transmission angles and transmission distance of the first subarray as parameters, which enables efficient construction of the overall steering vector for the WSA system. Building upon this novel steering vector, the analog precoder for each user can be easily constructed without complex computations such as singular value decomposition (SVD), and achieve acceptable performance of the sum SE.
  
  \item We analyze and compare the beam pattern and the sum SE performance of the MU-WSA and the compact array in both 2-D and 3-D cross-field communication systems. We show that the MU-WSA can provide additional distance domain resolution to distinguish users at the same angle, thus increasing the sum SE by over $60$\% at a power of $20$~dBm. Furthermore, extensive evaluation has been conducted to numerically investigate the performance of the proposed S-AO and SVR beamforming algorithms. The results show that the proposed S-AO algorithm in the SC system can achieve over 80\% of the sum (SE) of the FC system while reducing the number of phase shifters by $K^2$, thereby lowering power consumption. The SVR algorithm in the FC system can achieve over 95\% of the upper bound of SE but takes only 10\% of the running time of the SVD-based algorithms.
\end{itemize}

The remainder of this paper is organized as follows. In Sec.~\ref{section_Sys_channel_Model}, we present the channel model and the system model, based on which we formulate the beamforming design problem. Then, in Sec.~\ref{Sec_architecture_design}, we analyze the impact of the subarray spacing and the number of subarrays on the spectral efficiency, based on which the closed-form expression of the optimal subarray spacing is derived. In Sec.~\ref{section_HBF}, we propose the S-AO algorithm in the SC system and a low-complexity SVR algorithm in the FC system. Simulation results are shown in Sec.~\ref{section_peformance} where we first compare the beam pattern of the MU-WSA and compact array architecture in 2-D and 3-D scenarios, and we then evaluate the performance of the proposed algorithms. Conclusions are drawn in Sec.~\ref{section_conclusion}.

\textit{Notations:} $\mathbf{A}$ is a matrix, $\mathbf{a}$ is a vector and $a$ is a scalar; $\mathbf{A}(m,n)$ represents the element of $\mathbf{A}$ in row $m$ and column $n$; $\mathbf{I}$ is the identity matrix. $\operatorname{blkdiag}(\mathbf{A}_1, \cdots, \mathbf{A}_K)$ is the block diagonal matrix that composed of $\mathbf{A}_1$ to $\mathbf{A}_K$. $(\cdot)^T$ and $(\cdot)^H$ represents the transpose and conjugate transpose of a matrix. $| \cdot |$ and $\Vert \cdot \Vert_F$ denotes the modulus and the Frobenius norm. $\operatorname{det} (\cdot)$ represents the determinant. $\lceil a \rceil$ denotes the smallest integer no less than $a$. $\mathbb{C}^{m\times n}$ is the set of complex-valued matrices of dimension $m \times n$. 

\begin{figure*} %H为当前位置，!htb为忽略美学标准，htbp为浮动图形
\centering %图片居中
\includegraphics[width = 0.9\textwidth]{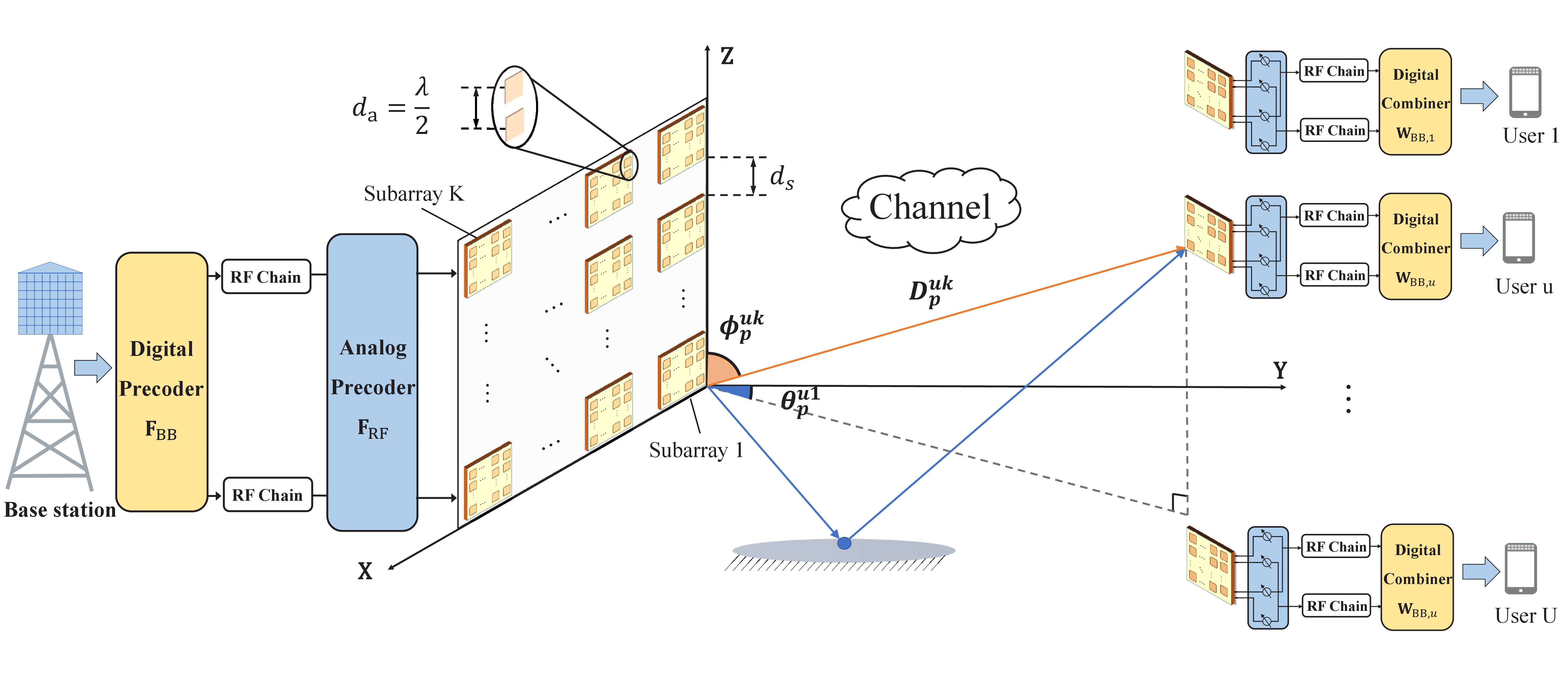} 
\caption{Block diagram of the MU-WSA system with hybrid beamforming architecture.} %最终文档中希望显示的图片标题
\label{Fig. model} %用于文内引用的标签
\end{figure*}

\section{System and Channel Models}
\label{section_Sys_channel_Model}
In this section, we introduce the channel model and the system model for the MU-WSA hybrid beamforming architecture.

\subsection{Cross Near-and-Far-Field (CNFF) Channel Model}
\label{section_Channel_Model}
As shown in Fig.~\ref{Fig. model}, we consider a downlink MU-WSA system where the BS serves $U$ users simultaneously. At each user, one compact UPA antenna array is equipped due to the limited size of the user equipment. The BS is equipped with $K$ widely-spaced subarrays on the X-Z plane. Within each subarray, the antennas are compactly spaced with half-wavelength. Therefore, due to the small size of the subarray, the Rayleigh distance between the subarray and each user is considered far less than the communication distance, resulting in the far-field propagation scheme. Therefore, according to the planar-wave assumption, the channel matrix between the $u^{\textrm{th}}$ user and the $k^{\textrm{th}}$ subarray at BS can be represented as~\cite{Spacespartial}
\begin{equation}
\label{equation_subchannel}
\mathbf{H}^{uk} = \sum_{p=1}^{N_{p}} \alpha_{p} \mathbf{a}_{r p}^{uk} \left( \phi_{r p}^{uk}, \theta_{r p}^{uk} \right) \mathbf{a}_{t p}^{uk}\left( \phi_{t p}^{uk}, \theta_{t p}^{uk}\right)^{H},
\end{equation}
where $N_p$ is the number of propagation paths for each user, i.e., $N_{p,u}=N_p, \forall u$. $\alpha_{p}$ denotes the complex path gain of the $p^{\textrm{th}}$ multipath. The vectors $\mathbf{a}_{r p}^{uk}\left( \phi_{r p}^{uk}, \theta_{r p}^{uk} \right) ^{uk}$ and $\mathbf{a}_{t p}^{uk}\left( \phi_{t p}^{uk}, \theta_{t p}^{uk}\right)$ are the receive and transmit array response vectors, where $\phi$ and $\theta$ refer to the azimuth and elevation angle, respectively. 

However, since the subarrays are widely-spaced, the effective array aperture of the whole array is increased, which in turn extends the Rayleigh distance and enlarges the near-field region. Consequently, the impact of the spherical wavefront must be taken into account across different subarrays, necessitating the use of the near-field model. In this context, the transmission and reception angles of the subarrays cannot be assumed to be identical, and the transmission distance of antennas in different subarrays must be calculated independently using the near-field SWM. Specifically, the channel between the first antenna of different subarrays and the first antenna at each user can be written as
\begin{subequations}
\label{eq:SWM}
\begin{align}
    \mathbf{H}_u^{\text{SWM}} &= \sum_{p=1}^{N_{p}} \begin{bmatrix}
        \alpha_{p}^{u1}, & \cdots, & \alpha_{p}^{uK}
    \end{bmatrix}\\
    & \approx \sum_{p=1}^{N_{p}} |\alpha_{p}|\begin{bmatrix}
        e^{-j\frac{2\pi}{\lambda}D_p^{u1}}, & \cdots, & e^{-j\frac{2\pi}{\lambda}D_p^{uK}}
    \end{bmatrix} \label{eq:temp},
\end{align}
\end{subequations}
where eq. (\ref{eq:temp}) is obtained since the complex path gain between two antennas $i$ and $l$ can be approximated by $\alpha_p^{il} \approx |\alpha_p^{11}|e^{-j\frac{2\pi}{\lambda}D_p^{il}}$~\cite{HSPM}.

As a result, the channel matrix of the $u^{\textrm{th}}$ user is constructed as a combination of the near-field SWM among different subarrays and the far-field PWM within each subarray, which can be written as~\cite{HSPM}
\begin{equation}
\begin{aligned}
    \mathbf{H}_u & = \begin{bmatrix} \mathbf{H}^{u1} & \cdots & \mathbf{H}^{uk}
    \end{bmatrix} \\
    & = \sum_{p=1}^{N_p}|\alpha_p| \left[
        e^{-j\frac{2\pi}{\lambda}D_p^{u1}}\mathbf{a}_{r p}^{u1}(\mathbf{a}_{t p}^{u1})^H, \cdots \right.\\
    & \quad \quad \quad \quad \quad \quad \left. ,\cdots, e^{-j\frac{2\pi}{\lambda}D_p^{uK}}\mathbf{a}_{r p}^{uk}(\mathbf{a}_{t p}^{uk})^H \right].
    \label{user_channel}
\end{aligned}
\end{equation}
Comparing with conventional PWM where $\operatorname{rank}(\mathbf{H}_u) = N_p$, the upper bound of the rank of $\mathbf{H}_u$ increases such that $N_p \leq \operatorname{rank}(\mathbf{H}_u) \leq KN_p$, for $K \leq \min \{\frac{N_t}{N_p}, \frac{N_r}{N_p}\}$.
While comparing to the SWM model where the channel response for each transmission and reception antenna should be calculated independently, the CNFF model requires fewer parameters, which leads to simpler calculation when designing the architecture in Sec.~\ref{Sec_architecture_design} and beamforming algorithms in Sec.~\ref{section_HBF}. The overall channel matrix can be written as $\mathbf{H}=[\mathbf{H}_1^T,\cdots,\mathbf{H}_U^T]^T$.

\subsection{System Model for MU-WSA}
\label{section_system_model}
At the receiver side, one compact UPA antenna array with $L_r$ RF chains is fully-connected to $N_r$ antennas to support the reception of $N_s \leq L_r$ data streams. While the BS is equipped with $L_t$ RF chains that can connect to the $N_t$ antennas either through FC or SC. To fully utilize the multiplexing gain of the MIMO system, we assume $UN_s = L_t = UL_r$ in this work.

At the BS, the transmitted symbols $\mathbf{s}=\left[\mathbf{s}_1^T, ..., \mathbf{s}_U^T\right]^T$ first go through a diagonal power allocation matrix $\mathbf{P} = \operatorname{blkdiag}\left(\mathbf{P}_1, ..., \mathbf{P}_U \right) \in \mathbb{C}^{UN_s\times UN_s}$, where $\mathbf{P}_u$ is the $u^{\textrm{th}}$ diagonal power allocation matrix. Let $\Vert \mathbf{P} \Vert^2_F = P_t$, where $P_t$ is the total power constraint. After power allocation, the symbols are processed by a baseband precoder $\mathbf{F}_{\mathrm{BB}}=[\mathbf{F}_{\mathrm{BB},1},\cdots, \mathbf{F}_{\mathrm{BB},U}] \in \mathbb{C}^{L_t \times UN_s}$ where $\mathbf{F}_{\mathrm{BB},u}$ is the digital precoder for the $u^{\mathrm{th}}$ user, followed by an RF precoder $\mathbf{F}_{\mathrm{RF}}  = [\mathbf{F}_1, \cdots, \mathbf{F}_K]$ where $ \mathbf{F}_k \in \mathbb{C}^{\frac{N_t}{K} \times L_t}$ represents the analog precoder of each subarray. At the $u^{\textrm{th}}$ user, the received signal is processed by the $u^{\textrm{th}}$ RF combiner $\mathbf{W}_{\mathrm{RF},u} \in \mathbb{C}^{N_r \times L_r}$, followed by the $u^{\textrm{th}}$ baseband combiner $\mathbf{W}_{\mathrm{BB},u} \in \mathbb{C}^{L_r \times N_s}$. Therefore the received signal of the $u^{\textrm{th}}$ user is described as
\begin{equation}
\label{Eq: signal}
\begin{aligned}
\tilde{\mathbf{y}}_{u}&=\underbrace{\mathbf{W}_{\mathrm{BB},u}^H \mathbf{W}_{\mathrm{RF},u}^H \mathbf{H}_{u} \mathbf{F}_{\mathrm{RF}} \mathbf{F}_{\mathrm{BB},u} \mathbf{P}_u \mathbf{s}_{u}}_{\text {desired signals }}\\
&+\underbrace{\mathbf{W}_{\mathrm{BB},u}^{H}\mathbf{W}_{\mathrm{RF}_u}^H \mathbf{H}_{u} \sum_{\ell \neq u} \mathbf{F}_{\mathrm{RF}} \mathbf{F}_{\mathrm{BB},{\ell}} \mathbf{P}_{\ell} \mathbf{s}_{\ell} }_{\text {effective interference}}\\
&+\underbrace{\mathbf{W}_{\mathrm{BB},u}^{H}\mathbf{W}_{\mathrm{RF},u}^H \mathbf{n}_{u}}_{\text {effective noise }},
\end{aligned}
\end{equation}
where $\mathbf{H}_u \in \mathbb{C}^{N_r \times N_t}$ denotes the channel matrix of the $u^{\textrm{th}}$ user. $\mathbf{n}_u \in \mathbb{C}^{N_r \times 1}$ represents the $u^{\textrm{th}}$ noise vector, following the complex Gaussian distribution as $\mathcal{CN}(0,\sigma_n^2)$.

Therefore, the achievable spectral efficiency of the $u^{\textrm{th}}$ user is represented as~\cite{HBD}
\begin{equation}
    \begin{aligned}
    SE_u = \log _{2} \operatorname{det}\left( \mathbf{I} \right. & + \left. \mathbf{R}_{u}^{-1} \mathbf{W}_{\mathrm{BB},{u}}^{H} \mathbf{W}_{\mathrm{RF},{u}}^{H} \mathbf{H}_{u} \mathbf{F}_{\mathrm{RF}} \mathbf{F}_{\mathrm{BB},{u}} \mathbf{P}_u \right.\\
& \left. \times \mathbf{P}_u^H \mathbf{F}_{\mathrm{BB},{u}}^{H} \mathbf{F}_{\mathrm{RF}}^{H} \mathbf{H}_{u}^{H} \mathbf{W}_{\mathrm{RF},{u}} \mathbf{W}_{\mathrm{BB},{u}}\right),
    \end{aligned}
    \label{eq:SE_u}
\end{equation}
where $\mathbf{R}_{u}$ denotes the covariance matrix of the user interference and the noise, i.e.,
\begin{equation}
\begin{aligned}
\mathbf{R}_{u} &=  \sum_{n \neq u}^{U} \mathbf{W}_{\mathrm{BB},{u}}^{H} \mathbf{W}_{\mathrm{RF},{u}}^{H} \mathbf{H}_{u} \mathbf{F}_{\mathrm{RF}} \mathbf{F}_{\mathrm{BB},{n}}\mathbf{P}_n \\ &\times \mathbf{P}_n^H \mathbf{F}_{\mathrm{BB},{n}}^{H} \mathbf{F}_{\mathrm{RF}}^{H} \mathbf{H}_{u}^{H} \mathbf{W}_{\mathrm{RF},{u}} \mathbf{W}_{\mathrm{BB},{u}}\\
&+\sigma_{\mathrm{n}}^{2} \mathbf{W}_{\mathrm{BB},{u}}^{H} \mathbf{W}_{\mathrm{RF},{u}}^{H} \mathbf{W}_{\mathrm{RF},{u}} \mathbf{W}_{\mathrm{BB},{u}}.
\end{aligned}
\end{equation}

\subsubsection{Fully-connected MU-WSA}
In a fully-connected WSA, each RF chain connects to all the antennas. Therefore, since the analog beamformers $\mathbf{F}_{\mathrm{RF}}$ and $\mathbf{W}_{\mathrm{RF},u}$ are implemented by phase shifters, they should satisfy the constant-modulus constraint, i.e., $\left|\mathbf{F}_{\mathrm{RF}}(m,n)\right| = \frac{1}{\sqrt{N_t}}$, and $\left|\mathbf{W}_{\mathrm{RF},{u}}(m, n)\right|=\frac{1}{\sqrt{N_r}}$. Additionally, to satisfy the transmitter’s power constraint, we normalize $\mathbf{F}_{\mathrm{BB}}$ such that $\left\|\mathbf{F}_{\mathrm{RF}} \mathbf{F}_{\mathrm{BB}}\right\|_{F}^{2}=UN_{s}$. 

\subsubsection{Sub-connected MU-WSA}
In the sub-connected WSA, we consider a more general scenario where each subarray is connected to multiple RF chains, in contrast to traditional SC architectures that assign one RF chain per subarray as stated in Sec.~\ref{section_intro}. In this context, the $L_t$ RF chains are divided uniformly into $K$ groups, with each group connected to one subarray. This results in a reduction of phase shifters by a factor of $K^2$ compared to FC systems. As a result, the analog precoder adopts a block-diagonal structure as
\begin{equation}
\mathbf{F}_{\mathrm{RF}}=\begin{bmatrix} \mathbf{F}_1 & 0 &\ldots & 0 \\ 0 & \mathbf{F}_2 & \ldots & 0 \\ \vdots& \vdots & \ddots & \vdots \\ 0 &\ldots & 0 &\mathbf{F}_K \end{bmatrix},
\label{eq: block_diagonal}
\end{equation}
where $\mathbf{F}_j \in \mathbb{C}^{\frac{N_t}{k} \times \frac{L_t}{k}}$ represents the matrix of the analog precoder for each subarray, and should satisfy $\left|\mathbf{F}_j(m,n)\right| = \frac{1}{\sqrt{N_t/K}}, \forall j$.

\subsection{Problem Formulation}
\label{section_problem}
In MU-WSA architecture, the subarray spacing $d_s$ and the number of subarrays $K$ need to be investigated since they have a significant impact on the system performance~\cite{InterIntra}. On one hand, as analyzed in Sec.~\ref{section_Channel_Model}, the rank of the channel matrix is directly related to $K$, thus affecting the SDoF and the sum SE. On the other hand, the values of $K$ and $d_s$ affect the positions of each antenna and each subarray, which further influence the transmission distance $D_p^{uk}$ and the array response vectors $\mathbf{a}_{tp}^{uk}$ and $\mathbf{a}_{rp}^{uk}$ of the channel matrix in eq. (\ref{user_channel}).

Specifically, by denoting the bottom right antenna as the reference antenna of each subarray, the position of the first subarray can be written as $p_1 = (0,0,h_t)$, where $h_t$ denotes the height of the BS. For simplicity, we assume the number of antennas in each subarray on the x- and z-axis are the same, which equals $n = \sqrt{\frac{N_t}{K}}$. Therefore, the distance between different subarrays on the x- and z-axis are also the same, which can be calculated as $d = \frac{(n-1)}{2}\lambda+d_s$. We denote the position of the $k^{\mathrm{th}}$ subarray as $p_k = (x_k,0,z_k)$. Then, based on the geometry shown in Fig.~\ref{Fig. model}, we can derive
\begin{equation}
\begin{aligned}
        x_k &= x_1+(k_x-1)d = (k_x-1)(\frac{(n-1)}{2}\lambda+d_s),\\
        z_k &= z_1+(k_z-1)d = h_t+(k_z-1)(\frac{(n-1)}{2}\lambda+d_s),
\end{aligned}
\end{equation}
where $k_x = k-(k_z-1)\sqrt{K}$ and $k_z=\lceil\frac{k}{\sqrt{K}}\rceil$ are the index positions of the $k^{\mathrm{th}}$ subarray on the x- and z-axis, respectively. Then, the transmission distance of the LoS path can be calculated as
\begin{equation}
\begin{aligned}
    & D_1^{uk} = \sqrt{(x_k-x_u)^2+(y_k-y_u)^2+(z_k-z_u)^2}\\
    & = \sqrt{((k_x-1)d-x_u)^2+y_u^2+(h_t+(k_z-1)d-z_u)^2},
    \label{eq:distance}
\end{aligned}
\end{equation}
where $p_u = (x_u,y_u,z_u)$ is the position of user $u$. Moreover, the azimuth and elevation angles for the LoS path are also influenced by $(K,d_s)$ as
\begin{subequations}
\label{Eq:WSA angle}
\begin{align}
     \cos(\phi_{t}^{uk}) & = \frac{z_u-z_k}{D_1^{uk}} = \frac{z_u-h_t-(k_z-1)d}{D_1^{uk}},\\
     \sin(\theta_{t}^{uk}) & = \frac{x_u-x_k}{D_1^{uk}\cos{\phi_t^{uk}}} = \frac{x_u-(k_x-1)d}{D_1^{uk}\cos{\phi_t^{uk}}}.
\end{align}%
\end{subequations}
For NLoS paths, the angles and the transmission distance can be calculated similarly. 

As the $(K,d_s)$ pair influences the channel matrix and affects the system performance, it requires careful optimization. However, due to the limited space in practice, the size of the array aperture at the BS is constrained to be smaller than a certain threshold, e.g., between $0.2$~m and $1.2$~m~\cite{InterIntra,Theory}. Considering a square UPA with array aperture $S_t$ representing the length of its diagonal, we have
\begin{equation}
        S_t=(\sqrt{\frac{N_t}{2}}-\sqrt{\frac{K}{2}})\lambda+(\sqrt{2K}-\sqrt{2})d_s.
\end{equation}
Therefore, the optimization of $(K,d_s)$ is constrained by the maximum array aperture such that they can not be arbitrarily large or small.

Apart from the design of $(K,d_s)$ pair, the hybrid beamformers also need to be optimized to maximize the sum SE for both FC and SC systems. In FC systems, the overall optimization problem can be formulated as
\begin{subequations}
        \begin{align}
            \underset{(K,d_s), \mathbf{F}_{\mathrm{RF}}, \mathbf{F}_{\mathrm{BB}}, \mathbf{W}_{\mathrm{RF}}, \mathbf{W}_{\mathrm{BB}},\mathbf{P}}{\operatorname{maximize}} & \ SE = \sum_{u=1}^U SE_u \label{R in Original}\\
            \text { subject to } \quad \quad \quad &
            \operatorname{Tr}\left(\mathbf{P}^H\mathbf{P}\right) = P_t, \\
            & \left|\mathbf{F}_{\mathrm{RF}}(m, n)\right|^{2}=\frac{1}{N_t}, \forall m, n, \label{eq:FRF_constraint_FC}\\
            & \left|\mathbf{W}_{\mathrm{RF},{u}}(m, n)\right|^{2}=\frac{1}{N_r}, \forall u, m, n,\\
            & S_t^{\text{min}} \leq S_t \leq S_t^{\text{max}}. \label{eq:aperture_constraint}
        \end{align}
    \label{Original problem}%
\end{subequations}
Constraint (\ref{eq:aperture_constraint}) denotes the array aperture constraint where $S_t^{\text{min}}$ refers to the case of a compact array. For SC, constraint (\ref{eq:FRF_constraint_FC}) is replaced by $|\mathbf{F}_j(m,n)|^2 = \frac{1}{N_t/K}$. This is a non-convex problem due to the constant-modulus constraints of the analog beamformers. Moreover, the effect of $(K,d_s)$ is implicit within the expression of the channel matrix, which further complicates the design problem and makes it intractable to solve directly. Therefore, we separate this problem into two sub-problems, for which we first design the MU-WSA architecture by analyzing the impact of the number of subarrays and the subarray spacing. After that, the channel matrix can be determined, based on which we design the hybrid beamformers to maximize the sum SE as the second sub-problem.

\section{Design of MU-WSA Architecture}
\label{Sec_architecture_design}
In this section, we focus on the optimization of the MU-WSA architecture, i.e., design of the $(K,d_s)$ pair, to maximize the upper bound of the sum SE, which can be achieved by assuming optimal precoding and combining~\cite{HBF,ZF}. That is to say, we assume each user transmits its own data streams without interference, such that the performance of each user can be analyzed independently. Considering equal power allocation, the optimal SE, i.e., capacity, for each user is written as
\begin{equation}
    C_u = SE_u^{\text{optimal}}=\sum_{i=1}^{\operatorname{rank}(\mathbf{H}_u)} \log _2\left(1+\frac{\rho}{\sigma_n^2} r_i^2(\mathbf{H}_u)\right),
\end{equation}
where $\rho = P_t/\operatorname{rank(\mathbf{H}_u)}$ denotes the allocated power for each data stream, and $r_i(\mathbf{H}_u)$ represents the $i^{\mathrm{th}}$ singular value of $\mathbf{H}_u$. Since the THz channel has the LoS-dominant feature~\cite{InterIntra}, we further simplify the optimization problem as to maximize the capacity of the LoS channel, i.e., $N_p = 1$. Based on (\ref{user_channel}), the LoS channel can be derived as
\begin{equation}
\begin{aligned}
    & \mathbf{H}_u^{LoS} = \begin{bmatrix} \mathbf{H}^{u1} & \cdots & \mathbf{H}^{uk}
    \end{bmatrix} \\
    & \resizebox{0.9\hsize}{!}{$= |\alpha_1| \left[
        e^{-j\frac{2\pi}{\lambda}D_1^{u1}}\mathbf{a}_{r 1}^{u1}(\mathbf{a}_{t 1}^{u1})^H, \cdots, e^{-j\frac{2\pi}{\lambda}D_1^{uK}}\mathbf{a}_{r 1}^{uk}(\mathbf{a}_{t 1}^{uk})^H \right].$}
\end{aligned}
\end{equation}%
Moreover, since the correlation of the array response vectors with different elevation angles and different azimuth angles are approximately zero when the number of antennas approaches infinity~\cite{Multiple-access,HSPM_IRS}, the rank of the LoS channel approximately equals $K$. To this end, the design problem for the MU-WSA architecture is formed as
\begin{subequations}
        \begin{align}
           & \underset{(K,d_s)}{\operatorname{maximize}} \ C_u^{\text{LoS}} =  \sum_{i=1}^{K} \log _2\left(1+\frac{\rho}{K\sigma_n^2}r_i^2(\mathbf{H}_u^{\text{LoS}})\right)\\
           & \text { subject to } S_{t}^{\text{min}}\leq S_t \leq S_{t}^{\text{max}}.
        \end{align}
\end{subequations}%
A gradient-descent method may be feasible to find an optimal solution for this problem in P2P scenarios~\cite{InterIntra}, however, it is impossible to change the array architecture for each user at the same time in multi-user scenarios. To tackle this problem, we present the following Theorem as
\begin{theorem}\label{theorem}
For a given number of subarrays $K$, the capacity of the LoS channel increases monotonically with the subarray distance $d_s$ when the transmission distance satisfies $D^{uk} \geq 2\sqrt{2}(\sqrt{N_r}-1) S_t, \forall u,k$. 
\end{theorem}

\textit{Proof:} We first apply Jensen's inequality and obtain
\begin{equation}
SE^{\text{LoS}}_{\text{optimal}} \leq K\log_2(1+\frac{\rho}{K^2\sigma_n^2}\sum_{i=1}^Kr_i^2(\mathbf{H}_u^{\text{LoS}})).
\end{equation}
Since $\sum_{i=1}^Kr_i^2(\mathbf{H}_u^{\text{LoS}}) = ||\mathbf{H}_u^{\text{LoS}}||_F^2 = N_tN_r\alpha$ is a constant, the maximization of the upper bound becomes to make the inequality holds, namely, the singular values are equal, i.e., $r_1 = r_2 = ... = r_K$. To calculate the singular values of channel $\mathbf{H}_u^{\text{LoS}}$, we define $\mathbf{H}_u^{\text{LoS}} = \mathbf{A}_r\mathbf{\Lambda}(\mathbf{A}_t)^H$ where
\begin{subequations}
        \begin{align}
            \mathbf{A}_r &= \begin{bmatrix}
            \mathbf{a}_r^{u1} & \mathbf{a}_r^{u2} & \cdots & \mathbf{a}_r^{uK}
            \end{bmatrix},\\
            \mathbf{A}_t &= \operatorname{blkdiag}(\mathbf{a}_t^{u1},\mathbf{a}_t^{u2},\cdots,\mathbf{a}_t^{uK}) \\
            & = \begin{bmatrix}
                \bar{\mathbf{a}}_t^{u1}& \bar{\mathbf{a}}_t^{u2}& \cdots, \bar{\mathbf{a}}_t^{uK}
            \end{bmatrix},\\
            \mathbf{\Lambda} &= \operatorname{blkdiag}(\alpha_{1},\alpha_{2},\cdots,\alpha_{K}).
        \end{align}
\end{subequations}
Therefore, we have
\begin{subequations}
\begin{align}
    \mathbf{H}_u^{\text{LoS}}(\mathbf{H}_u^{\text{LoS}})^H & = \mathbf{A}_r\mathbf{\Lambda}\mathbf{A}_t^H\mathbf{A}_t\mathbf{\Lambda}^H\mathbf{A}_r^H\\
    & = \frac{N_t}{K}\mathbf{A}_r\mathbf{\Lambda}\mathbf{\Lambda}^H\mathbf{A}_r^H\\
    & = |\alpha|^2\frac{N_t}{K}\mathbf{A}_r\mathbf{A}_r^H.
\end{align}
\end{subequations}
Since the elements along the diagonal of $\mathbf{A}_r\mathbf{A}_r^H$ equal $K$, we need to minimize the difference between $\mathbf{A}_r\mathbf{A}_r^H$ and $K\mathbf{I}$. Namely, the optimization problem is equivalent to minimizing the difference between 
\begin{equation}
    f = ||\mathbf{A}_r\mathbf{A}_r^H-K\mathbf{I}||_F^2.
\end{equation}
The $(i,j)^{th}, i \neq j$ element in $\bar{\mathbf{A}}_r = \mathbf{A}_r\mathbf{A}_r^H$ can be represented as 
\begin{equation}
    \bar{\mathbf{A}}_r(i,j) = \sum_{k=1}^Ke^{j\frac{2\pi}{\lambda}d_a[(j_1-i_1)\sin (\theta_t^{uk})\cos (\phi_t^{uk})+(j_2-i_2)\sin (\phi_t^{uk})]},
\end{equation} 
where $j_1,j_2, i_1,i_2 \in [0,\sqrt{Nr}]$ denotes the index on x- and z-axis of the $j^{th}$ and $i^{th}$ antenna. Then, the objective function $f$ can be decomposed as
\begin{equation}
    f = \sum_{i=1}^{N_r}\sum_{j=1}^{N_r}|\bar{\mathbf{A}}_r(i,j)|^2.
\end{equation}
Since f increases monotonically with $|\bar{\mathbf{A}}_r(i,j)|^2$, we can further transform the minimization of f as the minimization of function $|\bar{\mathbf{A}}_r(i,j)|^2, \forall i,j$. Define $\bar{\mathbf{A}}_r(i,j) = \sum_{k=1}^Ke^{j\beta_{k,i,j}}$. Then, based on eq. (\ref{Eq:WSA angle}), $\beta_{k,i,j}$ can be transferred into a function of $d_s$. Therefore, we have
\begin{equation}
\begin{aligned}
    g(d_s)& =|\bar{\mathbf{A}}_r(i,j)|^2 \\
    & = (\sum_{k=1}^K \cos(\beta_{k,i,j}))^2+(\sum_{k=1}^K \sin(\beta_{k,i,j}))^2\\
    & = K+2\sum_{1\leq k<m\leq K}\cos(\beta_{k,i,j}-\beta_{m,i,j}).
\end{aligned}
\end{equation}
To prove the monotonicity, we calculate the derivative of $g$ as 
\begin{equation}
g'(d_s) = 2\sum_{k<m}[-\sin(\beta_{k,i,j}-\beta_{m,i,j})\cdot (\frac{\partial (\beta_{k,i,j}-\beta_{m,i,j})}{\partial d_s})].
\end{equation}
Next, we prove that when $d_s$ becomes larger, the function $f$ monotonically decreases. This proof is presented in the Appendix.\hfill $\blacksquare$

\textbf{Remark 1:} In the proof, we impose a stricter condition than necessary by assuming that each summand $|\mathbf{A}_r(i,j)|$ in the function $f$ is monotonically increasing. As a result, the threshold of the transmission distance, i.e., $\tau = 2\sqrt{2}(\sqrt{N_r}-1) S_t^{\text{max}}$, is in fact a upper bound of the real threshold. That is to say, in practice, the distance can be smaller than $\tau$, while the monotonicity still holds.

\textbf{Remark 2:} Since the threshold $\tau$ is proportional to the square root of the number of the receiving antennas, the Theorem is primarily applicable to multi-user scenarios where the user equipment is considered to have much fewer antenna elements than the base station. In contrast, for P2P scenarios where the number of antennas is approximately the same at both ends, the spectral efficiency may exhibit significant fluctuations as $d_s$ changes.

\textbf{Remark 3:} The prerequisite of the Theorem, i.e., $\tau = 2\sqrt{2}(\sqrt{N_r}-1) S_t^{\text{max}} < D^{uk}$, imposes a constraint that the array aperture must not be too large compared to the transmission distance. Therefore, in DAS where the distance between different subarrays can achieve tens of meters such that the array aperture and the transmission distance are often comparable, this Theorem is not applicable.

\begin{figure} %H为当前位置，!htb为忽略美学标准，htbp为浮动图形
\centering %图片居中
\includegraphics[width = 0.4\textwidth]{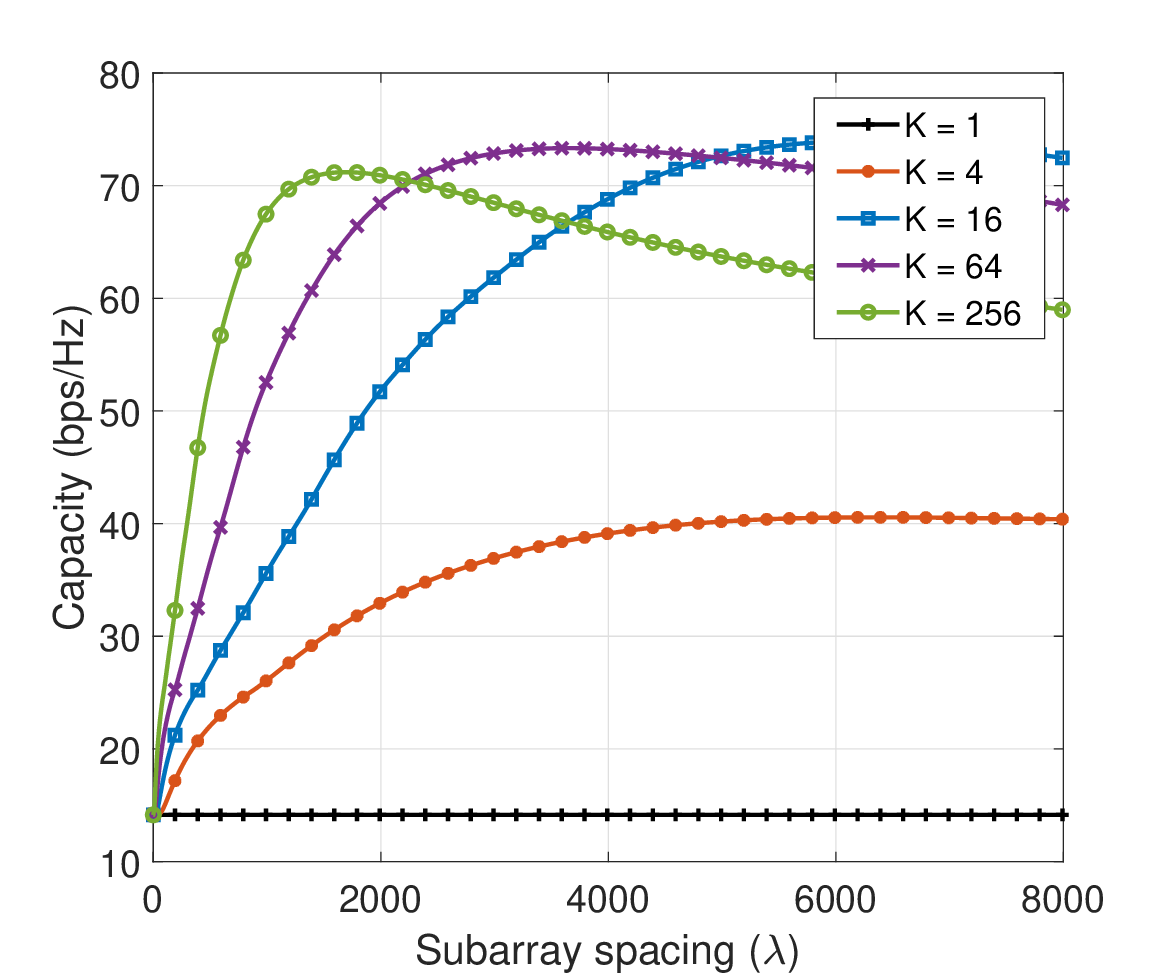} 
\caption{Impact of subarray spacing $d_s$ and the number of subarrays $k$ on the sum SE.} %最终文档中希望显示的图片标题
\label{Fig. ds_k_Impact} %用于文内引用的标签
\end{figure}
To directly illustrate the relationship between the channel capacity and the subarray spacing, we present the numerical analysis in Fig. \ref{Fig. ds_k_Impact}. The numbers of antennas at the BS and each user are set as $N_t = 1024$ and $N_r = 16$. We exhaustively sweep the choices of $K$ in $\mathcal{K} = [1, 4, 16,64,256]$. For each possible value of $K$, we vary the subarray spacing varies from $1 - 8000 \lambda$. Since the array aperture cannot be infinitely large, we omit the case for larger subarray spacing. In the figure, different colored lines represent different values of $K$, and each line shows the relationship between the capacity and the subarray spacing. At $K = 1$, the horizontal line is the baseline representing the compact array case, where the capacity is invariable with the change of the subarray spacing. For other values of $K$, the channel capacity first increases monotonically with the subarray spacing. This is because when $d_s$ increases, the difference of the angles $\phi_p^{(uk)}$ and $ \theta_p^{(uk)}$ in (\ref{equation_subchannel}) becomes larger, so that the difference between the sub-matrices $\mathbf{H}^{(uk)}$ is also larger, leading to a better spatial multiplexing gain. While as the subarray spacing continues to increase, eventually the prerequisite in Theorem 1 no longer holds, and the capacity begins to decrease. Note that when the number of subarrays is relatively small, i.e., $K=4$, the array aperture changes slowly with respect to the subarray spacing. As a result, the decrease in capacity is not immediately obvious. However, upon closer inspection of the figure, it is clear that the capacity still shows a declining trend from $40.6$~bps/Hz at $d_s = 6400\lambda$ to $40.3$~bps/Hz at $d_s = 8000\lambda$. This occurs because, when the array aperture becomes comparable to the transmission distance, the subarray spacing starts to have a more significant impact on the transmission distance. In other words, a larger subarray spacing leads to a larger transmission distance, which results in higher path loss and reduced channel capacity. However, in our MU-WSA systems, the subarray spacing is typically much smaller than the point at which the capacity starts to decrease. As a result, we can safely conclude that, in MU-WSA systems, the channel capacity increases with larger subarray spacing. 

This monotonicity is particularly suitable to a multi-user scenario. As the channel environment is typically dynamic, it is impractical to constantly adjust the positions of subarrays. Fortunately, based on the previous analysis, a positive conclusion can be drawn that as long as the number of subarrays and the array aperture are determined, the subarray spacing only needs to be set to the maximum, regardless of how the users change their positions. In practice, the number of subarrays and the array aperture are affected and determined by hardware constraint. For example, the number of subarrays should be smaller than the number of RF chains, and the available size for the antenna array is usually limited. Therefore, with fixed number of subarrays and maximum array aperture, the optimal subarray spacing can be easily obtained through
\begin{equation}
    d_s^{K,\text{max}} = \frac{\sqrt{2}S_{t}^{\text{max}}-\lambda(\sqrt{N_t}-\sqrt{K})}{2(\sqrt{K}-1)}.
    \label{eq:subarray_spacing}
\end{equation}
In this way, real-time adjustments of the architecture is unnecessary.

\section{Design of Hybrid Beamforming Algorithm}
\label{section_HBF}
After optimizing the array structure, the channel matrix can be determined for the design of the beamformers. To further simplify the beamforming sub-problem, we separately consider the digital and the analog stages. For the digital beamformers, the BD algorithm is applied to eliminate user interference. For the design of the analog beamformer, two different connection types are considered, i.e., the sub-connected and fully-connected WSA. For the sub-connected case, we propose an alternating-optimization-based algorithm, while for the FC architecture, we propose two low-complexity subarray-based algorithms.

\subsection{Digital Precoder and Combiner}
\label{Sec_BB}
For the digital precoder and combiner, we use the baseband block diagonalization method to mitigate the user interference, i.e., the effective interference in (\ref{Eq: signal}) should equal to zero. That is, the baseband precoder needs to satisfy $\bar{\mathbf{H}}_u \mathbf{F}_{\mathrm{BB},i} = 0$ for $i \neq u$, where $\bar{\mathbf{H}}_u = \mathbf{W}_{\mathrm{RF},u}^{H} \mathbf{H}_u \mathbf{F}_{\mathrm{RF}}$. By defining the interference channel matrix of the $u^{\textrm{th}}$ user as $\hat{\mathbf{H}}_u = \begin{bmatrix} \bar{\mathbf{H}}_1^T & \cdots & \bar{\mathbf{H}}_{u-1}^T & \bar{\mathbf{H}}_{u+1}^T &\cdots & \bar{\mathbf{H}}_U^T \end{bmatrix}^T$, the baseband precoder should lie in the null space of $\hat{\mathbf{H}}_u$. To start with, the singular value decomposition (SVD) of $\hat{\mathbf{H}}_u$ is
\begin{equation}
    \hat{\mathbf{H}}_u = \hat{\mathbf{U}}_u \hat{\mathbf{\Sigma}}_u \left[ \hat{\mathbf{V}}_{u}^{(1)}, \hat{\mathbf{V}}_u^{(0)} \right]^H,
\end{equation}
where $\hat{\mathbf{V}}_{u}^{(1)}$ is the first $(U-1)L_r$ right singular vectors of $\hat{\mathbf{H}}_u$, and $\hat{\mathbf{V}}_u^{(0)}$ denotes the orthogonal bases of the null space of $\hat{\mathbf{H}}_u$, i.e., $\bar{\mathbf{H}}_u \hat{\mathbf{V}}_i^{(0)} = 0, i \neq u$. As a result, we can eliminate the inter-user interference so that each user can transmit data stream via their own sub-channel $\bar{\mathbf{H}}_u \hat{\mathbf{V}}_u^{(0)}$. Without the inter-user interference, the sum SE (\ref{R in Original}) can then be maximized by achieving the maximum transmission SE of each user. Therefore, we further optimize the precoder and combiner on each sub-channel via SVD, i.e., $\bar{\mathbf{H}}_u \hat{\mathbf{V}}_u^{(0)} = \mathbf{U}_u \mathbf{\Sigma}_u \mathbf{V}_u^H$. Then, the expressions of the baseband precoder and combiner are given by
\begin{subequations}
    \label{Eq: baseband}
    \begin{align}
        & \mathbf{F}_{\mathrm{BB}} = \left[\hat{\mathbf{V}}_{1}^{(0)} \mathbf{V}_1^{\left(N_s\right)}, \cdots , \hat{\mathbf{V}}_{U}^{(0)} \mathbf{V}_U^{\left(N_s\right)} \right ],\\
        & \mathbf{W}_{\mathrm{BB},u} =\mathbf{U}_u^{\left(N_s\right)}, u \in\{1, \ldots, U\}.
    \end{align}%
\end{subequations}
where $\mathbf{U}_u^{\left(N_s\right)}$ and $\mathbf{V}_u^{\left(N_s\right)}$ are the first $N_s$ columns of $\mathbf{U}_u$ and  $\mathbf{V}_u$, respectively.

\subsection{Analog Precoder and Combiner}
After eliminating the inter-user interference, the sum SE can be written into a more compact form by defining the overall baseband and RF combiner as $\mathbf{W}_{\mathrm{BB}} = \operatorname{blkdiag} \left( \mathbf{W}_{\mathrm{BB},1}, \cdots, \mathbf{W}_{\mathrm{BB},U} \right)$ and $\mathbf{W}_{\mathrm{RF}} = \operatorname{blkdiag} \left( \mathbf{W}_{\mathrm{RF},1}, \cdots, \mathbf{W}_{\mathrm{RF},U} \right)$. As a result, SE can be expressed as
\begin{subequations}
    \begin{align}
        SE & = \sum_{u=1}^U SE_u\\
        \begin{split}
            & = \log _{2} \operatorname{det} \left(\mathbf{I} + \mathbf{R}^{-1} \mathbf{W}_{\mathrm{BB}}^H \mathbf{W}_{\mathrm{RF}}^{H} \mathbf{H F}_{\mathrm{RF}} \mathbf{F}_{\mathrm{BB}}\mathbf{P} \right.\\
    & \quad \quad \quad \quad \quad \quad \left. \times \mathbf{P}^H \mathbf{F}_{\mathrm{BB}}^H \mathbf{F}_{\mathrm{RF}}^{H} \mathbf{H}^{H} \mathbf{W}_{\mathrm{RF}} \mathbf{W}_{\mathrm{BB}} \right)
        \end{split}\\
    & = \log _{2} \operatorname{det}\left(\mathbf{I}+\frac{ \mathbf{W}_{\mathrm{RF}}^{H} \mathbf{H F}_{\mathrm{RF}} \mathbf{F}_{\mathrm{RF}}^{H} \mathbf{H}^{H} \mathbf{W}_{\mathrm{RF}}}{\sigma_{\mathrm{n}}^{2} \mathbf{W}_{\mathrm{RF}}^{H} \mathbf{W}_{\mathrm{RF}}}\right),
    \end{align}%
\end{subequations}
As in UM-MIMO systems, the optimal analog combiners typically satisfy $\mathbf{W}_{\mathrm{RF}}^H\mathbf{W}_{\mathrm{RF}}\approx \mathbf{I}$~\cite{HDA,Spacespartial}, the objective function can be simplified as 
\begin{subequations}
\begin{align}
SE & \approx \log _{2} \operatorname{det}\left(\mathbf{I}+ \sigma_{\mathrm{n}}^{-2} \mathbf{W}_{\mathrm{RF}}^{H}  \mathbf{H}  \mathbf{F}_{\mathrm{RF}} \mathbf{F}_{\mathrm{RF}}^{H} \mathbf{H}^{H} \mathbf{W}_{\mathrm{RF}}   \right)
    \label{objectivefunction_WRF}\\
    & = \log _{2} \operatorname{det}\left(\mathbf{I}+ \sigma_{\mathrm{n}}^{-2} \mathbf{F}_{\mathrm{RF}}^{H} \mathbf{H}^{H} \mathbf{W}_{\mathrm{RF}} \mathbf{W}_{\mathrm{RF}}^{H} \mathbf{H}  \mathbf{F}_{\mathrm{RF}} \right). \label{objectivefunction_FRF}
\end{align}
\end{subequations}

\subsubsection{Alternating-Optimization-based Analog Beamforming in Sub-connected MU-WSA}
In the sub-connected MU-WSA system where the special block-diagonal structure of the analog precoder should be considered as in (\ref{eq: block_diagonal}). Since we consider a general case where each subarray can be connected to more than one RF chain, each block in $\mathbf{F}_{\mathrm{RF}}$ can be a matrix. To solve this problem, we propose to isolate the contribution of each sub-matrix and optimize one sub-matrix per iteration with other sub-matrices fixed.

We first consider the design of the analog precoder assuming fixed $\mathbf{W}_{\mathrm{RF}}$. Then, the matrix in (\ref{objectivefunction_FRF}) can be decomposed into four parts, i.e.,
\begin{equation}
    \begin{aligned}
& \log _{2} \operatorname{det}\left(\mathbf{I}+ \sigma_{\mathrm{n}}^{-2} \mathbf{F}_{\mathrm{RF}}^{H} \mathbf{A} \mathbf{F}_{\mathrm{RF}}\right) \\
=& \log _{2} \operatorname{det}\left(\begin{array}{cc}
\mathbf{I}+\frac{ \mathbf{F}_{j}^{H} \mathbf{A}_{jj} \mathbf{F}_{j}}{\sigma_{\mathrm{n}}^{2}} & \frac{ \mathbf{F}_{j}^{H} \mathbf{A}_{j:} \left(\mathbf{F}_{\mathrm{RF}}\right)_{-j}}{\sigma_{\mathrm{n}}^{2}} \\
\frac{\left(\mathbf{F}_{\mathrm{RF}}\right)_{-j}^{H} \mathbf{A}_{:j} \mathbf{F}_{j}}{ \sigma_{\mathrm{n}}^{2}} & \mathbf{I}+\frac{\left(\mathbf{F}_{\mathrm{RF}}\right)_{-j}^{H} (\mathbf{A})_{-jj}\left(\mathbf{F}_{\mathrm{RF}}\right)_{-j}}{ \sigma_{\mathrm{n}}^{2}}
\end{array}\right) \\
\overset{(a)}{=}& \log _{2} \operatorname{det}\left(\mathbf{B}_{j}\right)+\log _{2} \operatorname{det}\left(\mathbf{I}+\frac{1}{ \sigma_{\mathrm{n}}^{2}} \mathbf{F}_{j}^{H} \mathbf{X}_{j} \mathbf{F}_{j}\right),
\end{aligned}
\label{Formula: ananolg precoder}
\end{equation}
where
\begin{subequations}
\begin{align}
\mathbf{A}&=\mathbf{H}^{H} \mathbf{W}_{\mathrm{RF}} \mathbf{W}_{\mathrm{RF}}^{H} \mathbf{H} \notag \\
&= \begin{bmatrix}
\mathbf{A}_{11} & \mathbf{A}_{12} & \ldots & \mathbf{A}_{1k}\\ \mathbf{A}_{21} & \mathbf{A}_{22} & \ldots & \mathbf{A}_{2k}\\ \ldots & \ldots & \ldots  & \ldots \\ \mathbf{A}_{k1} & \mathbf{A}_{k2} & \ldots & \mathbf{A}_{kk}
\end{bmatrix},\ \mathbf{A}_{ij} \in \mathbb{C}^{\frac{N_t}{k} \times \frac{N_t}{k}}, \\
\mathbf{B}_{j} &=\mathbf{I}+\frac{\left(\mathbf{F}_{\mathrm{RF}}\right)_{-j}^{H} (\mathbf{A})_{-jj}\left(\mathbf{F}_{\mathrm{RF}}\right)_{-j}}{ \sigma_{\mathrm{n}}^{2}}, \label{eq: Bj}\\
\mathbf{X}_{j}&=\mathbf{A}_{jj}-\frac{1}{ \sigma_{\mathrm{n}}^{2}} \mathbf{A}_{j:}\left(\mathbf{F}_{\mathrm{RF}}\right)_{-j} \mathbf{B}_{j}^{-1}\left(\mathbf{F}_{\mathrm{RF}}\right)_{-j}^{H} \mathbf{A}_{:j}, \label{eq: Xj}
\end{align}
\end{subequations}
where $\mathbf{(A)}_{-jj}$ is formed by deleting the rows and columns of $\mathbf{A}$ where the submatrix $\mathbf{A}_{jj}$ is located. $\mathbf{A}_{:j} = \left[\mathbf{A}_{1j}^T, \cdots, \mathbf{A}_{kj}^T\right]^T$, and $ \mathbf{A}_{j:} = \left[ \mathbf{A}_{j1}, \cdots, \mathbf{A}_{jk}\right]$. Moreover, $(\mathbf{F}_{\mathrm{RF}})_{-j}$ is the matrix $\mathbf{F}_{\mathrm{RF}}$ except the $j^{\textrm{th}}$ submatrix $\mathbf{F}_j$. Step $(a)$ in (\ref{Formula: ananolg precoder}) holds due to the theory that for any block matrix with invertible submatrix $\mathbf{D}$, we have 
\begin{equation}
    \operatorname{det} \left(\begin{bmatrix}
\mathbf{A} & \mathbf{B} \\
\mathbf{C} & \mathbf{D}
\end{bmatrix}\right)=\operatorname{det}\left(\mathbf{D}\right) \operatorname{det}\left(\mathbf{A}-\mathbf{B} \mathbf{D}^{-1} \mathbf{C}\right).
\end{equation} 
Since $\mathbf{B}_j$ in (\ref{eq: Bj}) is not related to $\mathbf{F}_j$, we only need to consider the second term in (\ref{Formula: ananolg precoder}). The optimal solution to the unconstrained $\mathbf{F}_j$ is then defined by the first $\ell_t = L_t/k$ columns of the right singular vectors of $\mathbf{X}_j= \mathbf{U}_{\mathbf{X}_j}\mathbf{\Sigma}_{\mathbf{X}_j} \mathbf{V}_{\mathbf{X}_j}^H$, i.e., $\mathbf{F}_j = \mathbf{V}_{\mathbf{X}_j}(:,1:\ell_t)$.

Next, we fix $\mathbf{F}_{\mathrm{RF}}$ and extract the contribution of the analog combiner $\mathbf{W}_{\mathrm{RF}}$ to the sum SE (\ref{objectivefunction_WRF}). Since the spectral efficiency of each user is only related to its own combiner, we design the analog combiner for each user individually to maximize the overall spectral efficiency, which can be given as
\begin{equation}
\begin{aligned}
    & \log _{2} \operatorname{det}\left(\mathbf{I}+ \sigma_{\mathrm{n}}^{-2} \mathbf{W}_{\mathrm{RF}}^{H} \mathbf{D} \mathbf{W}_{\mathrm{RF}}\right)\\
    = & \sum_{u=1}^U \log _{2} \operatorname{det}\left(\mathbf{I}+ \sigma_{\mathrm{n}}^{-2} \mathbf{W}_{\mathrm{RF},u}^{H} \mathbf{D}_u \mathbf{W}_{\mathrm{RF},u}\right),
\end{aligned}
\end{equation}
where $\mathbf{D}=\mathbf{H}^{H} \mathbf{F}_{\mathrm{RF}} \mathbf{F}_{\mathrm{RF}}^{H} \mathbf{H}$ and $\mathbf{D}_u = \mathbf{H}_u^{H} \mathbf{F}_{\mathrm{RF}} \mathbf{F}_{\mathrm{RF}}^{H} \mathbf{H}_u$. Then, the optimal solution of the unconstrained $u^{\textrm{th}}$ analog combiner is given by the first $L_r$ columns of the right singular vectors of $\mathbf{D}_u= \mathbf{U}_{\mathbf{D}_u}\mathbf{\Sigma}_{\mathbf{D}_u} \mathbf{V}_{\mathbf{D}_u}^H$, i.e., $\mathbf{W}_{\mathrm{RF},u} = \mathbf{V}_{\mathbf{D}_u}(:,1:L_r)$. 

After finding the unconstrained analog beamformers, we can derive the corresponding solution in the constrained case. 
According to~\cite{HBF}, the elements in the optimal constrained precoder should share the phase of the corresponding element in the unconstrained case. That is, the constrained analog precoder is given by $\mathbf{F}_j = \frac{1}{\sqrt{N_t}} e ^{i \angle{\mathbf{V}_{\mathbf{X}_j}(:,1:\ell_t)}}$. Similarly, the optimal constrained solution for $\mathbf{W}_{\mathrm{RF},u}$ is given by $\mathbf{W}_{\mathrm{RF},u} = \frac{1}{\sqrt{N_r}} e ^{i \angle{\mathbf{V}_{\mathbf{D}_u}(:,1:L_r)}}$.

A summary of the overall proposed algorithm is presented in Algorithm \ref{Ag: 1}. Specifically, we first optimize the $k$ sub-matrices in $\mathbf{F}_{\mathrm{RF}}$, and then iterate between $\mathbf{F}_{\mathrm{RF}}$ and $\mathbf{W}_{\mathrm{RF}}$ till a stopping criterion triggers. Next, with fixed analog beamforming matrices, we perform a BD method to obtain the baseband precoder and combiner. Finally, we perform the water-filling algorithm for multi-users to obtain the power allocation matrix.
\begin{algorithm}
	\renewcommand{\algorithmicrequire}{\textbf{Input:}}
	\renewcommand{\algorithmicensure}{\textbf{Output:}}
	\caption{Hybrid	beamforming with subarray-based alternating optimization (S-AO) in THz multi-user WSA systems}
	\label{alg1}
	\begin{algorithmic}[1]
		\REQUIRE $\mathbf{H}$, $K$
		\STATE Initialize $\mathbf{W}_{\mathrm{RF}}=\operatorname{blk} \operatorname{dig}\left(\mathbf{W}_{\mathrm{RF},1}, \ldots, \mathbf{W}_{\mathrm{RF},U}\right)$ where $\mathbf{W}_{\mathrm{RF},i} \in \mathbb{C}^{N_{r} \times L_r}$, and $\mathbf{F}_{\mathrm{RF}}=\left[\mathbf{F}_{1}, \ldots, \mathbf{F}_{K}\right]$ where $\mathbf{F}_{\mathrm{i}}=\mathbb{C}^{N_t \times \frac{L_t}{K}}$.
		\REPEAT
		\STATE Compute $\mathbf{A}=\mathbf{H}^{H} \mathbf{W}_{\mathrm{RF}} \mathbf{W}_{\mathrm{RF}}^{H} \mathbf{H}$
		\FOR {$j = 1$ to $K$}
		\STATE Compute $\mathbf{B}_j$ through (\ref{eq: Bj}) and $\mathbf{X}_j$ through (\ref{eq: Xj})
		\STATE $\mathbf{F}_j = \mathbf{V}_{\mathbf{X}_j}^{(l_t)}$ is the first $l_t$ columns of $\mathbf{V}_{\mathbf{X}_j}$
		\ENDFOR 
		\FOR {$u = 1$ to $U$}
		\STATE $\mathbf{D}_u = \mathbf{H}_u \mathbf{F}_{\mathrm{RF}} \mathbf{F}_{\mathrm{RF}}^H \mathbf{H}_u$
		\STATE $\mathbf{W}_{\mathrm{RF},u} =  \mathbf{V}_{\mathbf{D}_u}^{(L_r)}$ where $\mathbf{V}_{\mathbf{D}_u}^{(L_r)}$ is the first $L_r$ columns of $\mathbf{V}_{\mathbf{D}_u}$
		\ENDFOR
		\UNTIL The stopping criterion trigger is determined by the iteration number N, which will be discussed in the next section.
		\STATE Compute $\mathbf{F}_j = \frac{1}{\sqrt{N_t}} e ^{i \angle{\mathbf{F}_j}}$ and $\mathbf{W}_{\mathrm{RF},u} = \frac{1}{\sqrt{N_r}} e ^{i \angle{\mathbf{W}_{\mathrm{RF},u}}}$.
		\STATE Apply the baseband block diagonalization algorithm to calculate $\mathbf{F}_{\mathrm{BB}}$ and $\mathbf{W}_{\mathrm{BB},u}$.
		\STATE Normalize each column of $\mathbf{F}_{\mathrm{BB}}$ such that $\mathbf{F}_{\mathrm{BB}}(:,i) = \frac{\mathbf{F}_{\mathrm{BB}}(:,i)}{\left\|\mathbf{F}_{\mathrm{RF}} \mathbf{F}_{\mathrm{BB}}(:, i)\right\|_{F}}, i \in\left\{1, \ldots, L_t \right\}$.
		\STATE Compute $\mathbf{P}$ based on the water-filling algorithm.
		\ENSURE $\mathbf{F}_{\mathrm{RF}}, \mathbf{F}_{\mathrm{BB}}, (\mathbf{W}_{\mathrm{RF},u}, \mathbf{W}_{\mathrm{BB},u})_{u=1:U}, \mathbf{P}$
	\end{algorithmic} 
 \label{Ag: 1}
\end{algorithm}

\subsubsection{Low-complexity Subarray-based Analog Beamforming in Fully-connected MU-WSA}
Now we consider the fully-connected MU-WSA architecture where each subarray is connected to all the RF chains. In this case, we assume each user is equipped with one RF chain to receive $N_s=1$ data stream. Then, to reduce the complexity of the analog beamforming, we propose a subarray-based steering vector reconstruction (SVR) scheme, which avoids the utilization of the high-complexity operation of SVD. Since the channel between each subarray and the users can be represented by the far-field PWM, we first simplify the analog beamforming design by designing the analog precoder for each subarray. We denote the overall channel matrix as $\mathbf{H} = [\mathbf{H}_{sub}^1,\cdots,\mathbf{H}_{sub}^K]$, where $\mathbf{H}_{sub}^k$ represents the channel matrix between the $k^{\mathrm{th}}$ subarray and all the users. Then, based on eq.~(\ref{objectivefunction_FRF}), we have
\begin{equation}
\begin{aligned}
    SE & \approx \sum_{k=1}^K \log _{2} \operatorname{det}\left(\mathbf{I}+ \sigma_{\mathrm{n}}^{-2} \mathbf{F}_{\mathrm{RF},k}^{H} (\mathbf{H}_{sub}^k)^{H} \mathbf{W}_{\mathrm{RF}} \right.\\
    & \left. \quad \quad \quad \quad \times \mathbf{W}_{\mathrm{RF}}^{H} \mathbf{H}_{sub}^k  \mathbf{F}_{\mathrm{RF},k} \right)\\
    & = \sum_{k=1}^K SE_k
\end{aligned}
\end{equation}
Since $SE_k$ is independent of each other assuming fixed analog combiner, we can design the precoder $\mathbf{F}_{\mathrm{RF},k}$ in parallel. Specifically, the analog precoder can be obtained through the beam steering vector where the steering vector between user $u$ and the first subarray is represented as 
\begin{equation}
    \mathbf{a}_t^{u1} = [1, \cdots, e^{j\frac{2\pi}{\lambda}d_a ((N_L-1)\sin\theta_{t}^{u1}\cos(\phi_{t}^{u1})+(N_W-1) \sin (\phi_{t}^{u1})}]^T,
\end{equation}
where $N_L$ and $N_W$ are the number of antennas in a subarray at the x-axis and z-axis respectively. Then, the steering vector of the $k^{\mathrm{th}}$ subarray can be constructed by considering the phase difference between different subarrays, i.e.,
\begin{equation}
\begin{aligned}
    \mathbf{a}_t^{uk} & = e^{j\frac{2\pi}{\lambda}(D^{uk}-D^{u1})}\\
    \times & [1, \cdots, e^{j\frac{2\pi}{\lambda}d_a ((N_L-1)\sin\theta_{t}^{uk}\cos(\phi_{t}^{uk})+(N_W-1) \sin (\phi_{t}^{uk})}].
\end{aligned}
\end{equation}

Denote $\delta d_x^k = (k_x-1)d$ and $\delta d_z^k = (k_z-1)d$. Then, based on equation (\ref{eq:distance}), and applying the second order Taylor expansion that $\sqrt{1+\psi}\approx 1+ \frac{1}{2}\psi-\frac{1}{8}\psi^2$, we can derive 
\begin{subequations}
\begin{align}
    & \resizebox{1\hsize}{!}{$D^{uk} = \sqrt{(D^{u1})^2+2\delta d_x^k(x_1-x_u)+2\delta d_z^k(z_1-z_u)+(\delta d_x^k)^2+(\delta d_z^k)^2}$}\\
    &\quad \  = D^{u1}\sqrt{1+\psi}\\
    & \quad\ \approx D^{u1}+\frac{1}{2}\psi^{uk}D^{u1}-\frac{1}{8}(\psi^{uk})^2D^{u1}, \label{eq:Taylor}
\end{align}
\label{Eq:distance_tay}
\end{subequations}
where 
\begin{subequations}
\begin{align}
    \psi & = \frac{2\delta d_x^k(x_1-x_u)+2\delta d_z^k(z_1-z_u)+(\delta d_x^k)^2+(\delta d_z^k)^2}{(D^{u1})^2}\\
    \begin{split}
    & = 2[k_z-1-(k_x-1)\sin(\theta_t^{u1})]\cos(\phi_t^{u1})\frac{d}{D^{u1}}\\
    & \quad \quad \quad +[(k_x-1)^2+(k_z-1)^2](\frac{d}{D^{u1}})^2.
    \end{split}\label{eq:angle_distance}
    \end{align}
\end{subequations}

Therefore, the distance difference can be calculated as
\begin{equation}
    \Delta_D^{uk} = D^{uk}-D^{u1}=\frac{1}{2}\psi^{uk}D^{u1}-\frac{1}{8}(\psi^{uk})^2D^{u1}.
\end{equation}
As a result, when combining the beam steering vector for each subarray, the overall beam steering vector toward each user is formulated as
\begin{equation}
    \mathbf{a}_t^{(u)} = \begin{bmatrix}
        \mathbf{a}_t^{u1}\\ \vdots \\ \mathbf{a}_t^{uk}
    \end{bmatrix}.
    \label{Eq:WSA steering}
\end{equation}
Note that when the users are at the far-field region of MU-WSA, such that $D^{uk} \approx D, \forall k$, the reconstructed WSA steering vector is approximately the same as the traditional far-field beam steering vector. Finally, we obtain the resulting analog precoder as
\begin{equation}
    \mathbf{F}_{\mathrm{RF}} = [\mathbf{a}_t^{(1)}, \cdots, \mathbf{a}_t^{(U)}].
\end{equation}
The algorithm is summarized in Algorithm~\ref{alg:SVR}.

\begin{algorithm}
	\renewcommand{\algorithmicrequire}{\textbf{Input:}}
	\renewcommand{\algorithmicensure}{\textbf{Output:}}
	\caption{Analog beamforming based on steering vector reconstruction (SVR)}
	\begin{algorithmic}[1]
		\REQUIRE Target angle of the first subarray $\{\phi^{u1}\}$, $\{\theta^{u1}\}$, transmission distance $\{D^{u1}\}$
		\STATE Initialize $\mathbf{F}_{\mathrm{RF}} = [\ ]$.
        \FOR {$u = 1$ to $U$}	
        \FOR {$k = 1$ to $K$}
        \STATE Calculate $D^{uk}$ through (\ref{Eq:distance_tay})
        \STATE Calculate $\theta^{uk}$ and $\phi^{uk}$ through (\ref{Eq:WSA angle}).
        \ENDFOR
        \STATE Construct $\mathbf{a}_t^{(u)}$ through (\ref{Eq:WSA steering}).
        \STATE $\mathbf{F}_{\mathrm{RF}} = [\mathbf{F}_{\mathrm{RF}}|\mathbf{a}_t^{(u)}]$
        \ENDFOR
        \ENSURE $\mathbf{F}_{\mathrm{RF}}$
	\end{algorithmic}
 \label{alg:SVR}
\end{algorithm}

\section{Performance Evaluation}
\label{section_peformance}
In this section, we compare the performance of the MU-WSA and the compact array for both 2-D and 3-D cases. We then evaluate the performance of the proposed algorithm in the MU-WSA architecture and compare it with the state-of-the-art algorithms in the compact architecture under different user scenarios. The simulation setup is summarized in Table~\ref{Table: Simulation} where we set the maximum array aperture as $S_t^{max}=1$ m, and fix the number of subarrays as $K=4$. Then the subarray spacing $d_s = 0.692$~m can be obtained through eq.~(\ref{eq:subarray_spacing}).
\begin{table}
\centering
\caption{Simulation Setup.}
\label{Table: Simulation}%添加标题 设置标签
\begin{tabularx}{\linewidth}{|>{\centering\arraybackslash}p{6cm}|>{\centering\arraybackslash}X|}
\hline
\textbf{Parameter}& \textbf{Value}\\
\hline
Central frequency $f_c$ & $300$~GHz\\

Bandwidth B & $1$~GHz\\

Number of multipath components $N_p$ & 2\\

Number of transmit antennas $N_t$ & 1024\\

Number of receive antennas $N_r$ & 16\\

Maximum array aperture $S_t^{\text{max}}$ & $1$~m\\
2D scenario: BS, UE at height $h_t=h_r$ & $20$~m\\

3D scenario: BS, UE at height $h_t$, $h_r$ & $20$~m, $1.5$~m\\

Number of data streams for each user $N_s$ & 1\\

Number of users $U$ & 20\\

Number of transmit and receive RF chains $L_t$, $L_r$ & 20, 1\\

Number of subarrays at x- and z-axis of MU-WSA & 2, 2\\

Subarray spacing $d_s$ & 692$\lambda$ = 0.692~m\\
\hline
\end{tabularx}
\end{table}

\subsection{Cross Field Analysis of MU-WSA and Compact Array}
\label{Cross-field}
In this section, we analyze the beam pattern and the sum spectral efficiency for compact and MU-WSA architecture in the cross near-and-far-field, considering both 2-D and 3-D scenarios.

\subsubsection{2-D Cross Field Analysis}
\begin{figure*}
\begin{minipage}{0.49\textwidth}
\subfigure[Normalized beamforming gain in compact array.] {
\label{fig_2D:a}
\includegraphics[width=0.48\textwidth, height=0.45\textwidth]{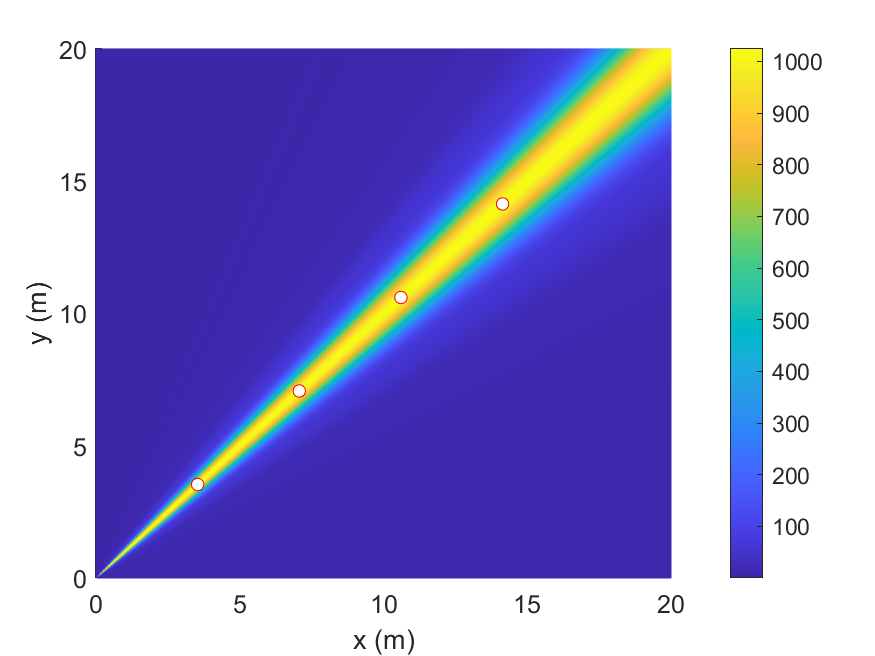}}\hfill
\subfigure[Normalized beamforming gain in MU-WSA.]{
\label{fig_2D:b}     
\includegraphics[width=0.48\textwidth, height=0.45\textwidth]{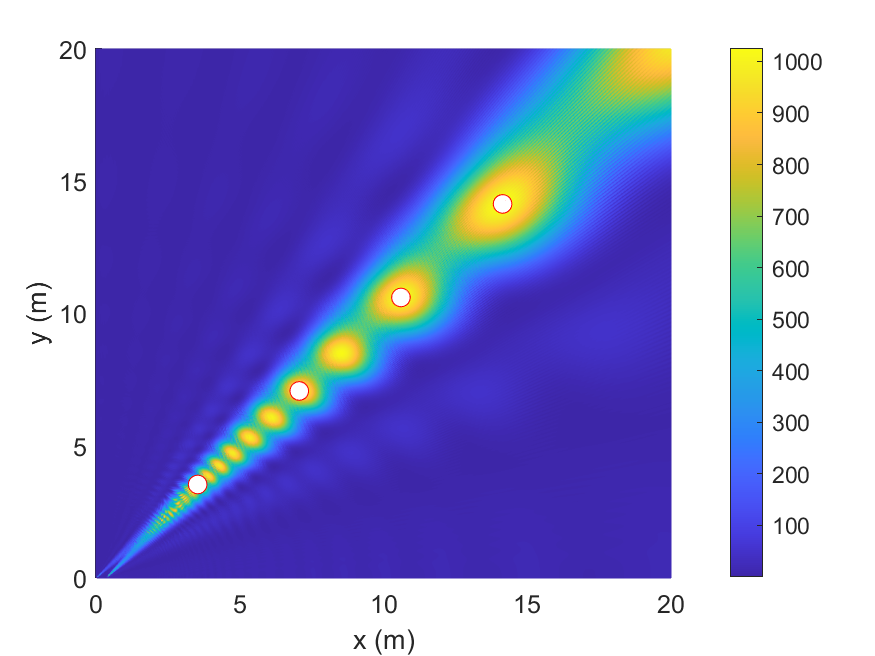}} 
\caption{Beamforming gain for 2D users at the same angle.}
\label{fig:2D-pattern}
\end{minipage}\hfill
\begin{minipage}{0.5\textwidth}
\subfigure[Users at the same azimuth angle.] {
\label{fig_2D_SE:a}
\includegraphics[width=0.44\textwidth, height=0.43\textwidth]{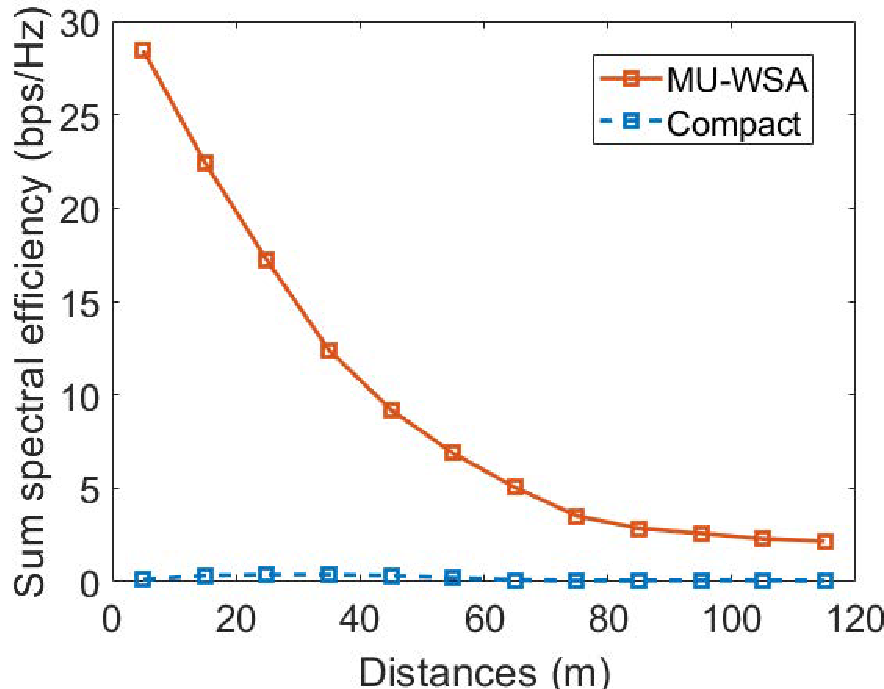}}\hfill
\subfigure[Users at different azimuth angles.]{
\label{fig_2D_SE:b} 
\includegraphics[width=0.48\textwidth, height=0.45\textwidth]{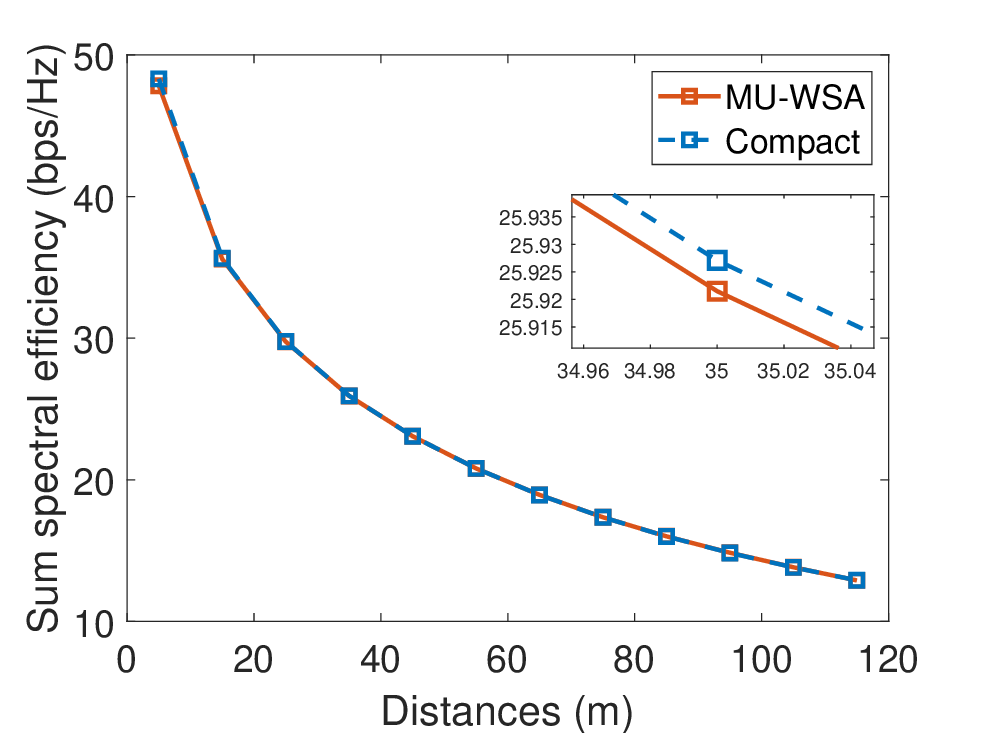}}
\caption{2D scenario: optimal sum SE versus transmission distance.}
\label{fig:2D_SE}
\end{minipage}

\begin{minipage}{0.49\textwidth}
\subfigure[Normalized beamforming gain in compact array.] {
\label{fig_3D:a}
\includegraphics[width=0.48\textwidth, height=0.45\textwidth]{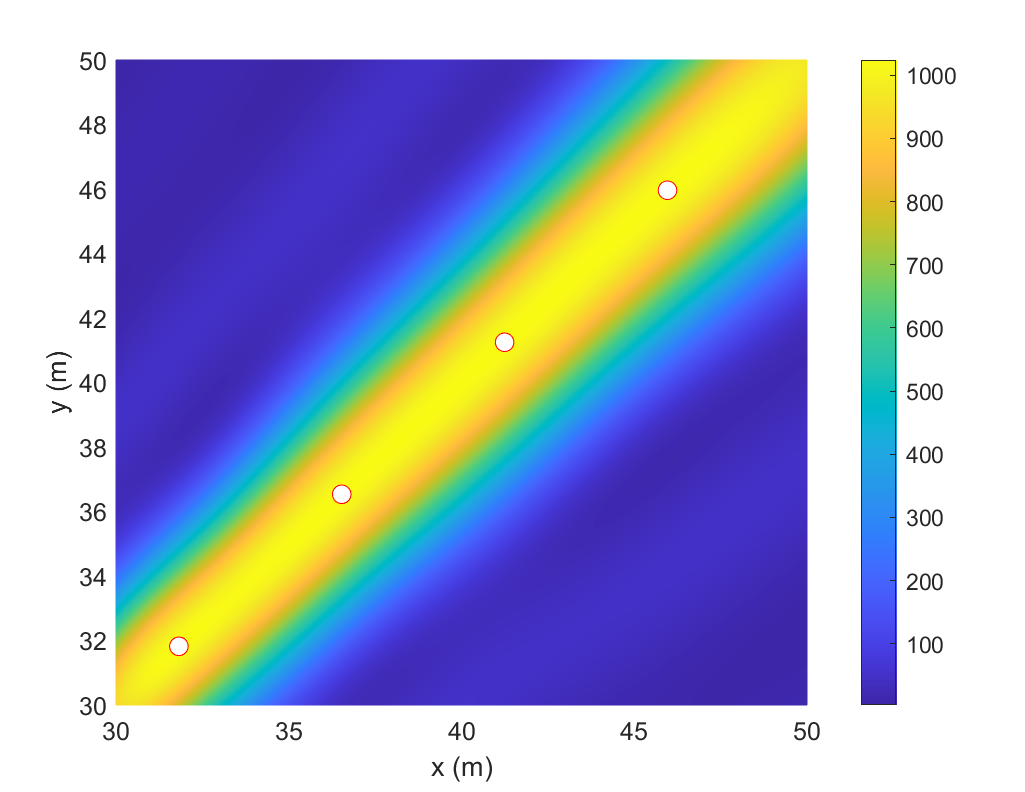}}\hfill
\subfigure[Normalized beamforming gain in MU-WSA.]{
\label{fig_3D:b}   
\includegraphics[width=0.48\textwidth, height=0.45\textwidth]{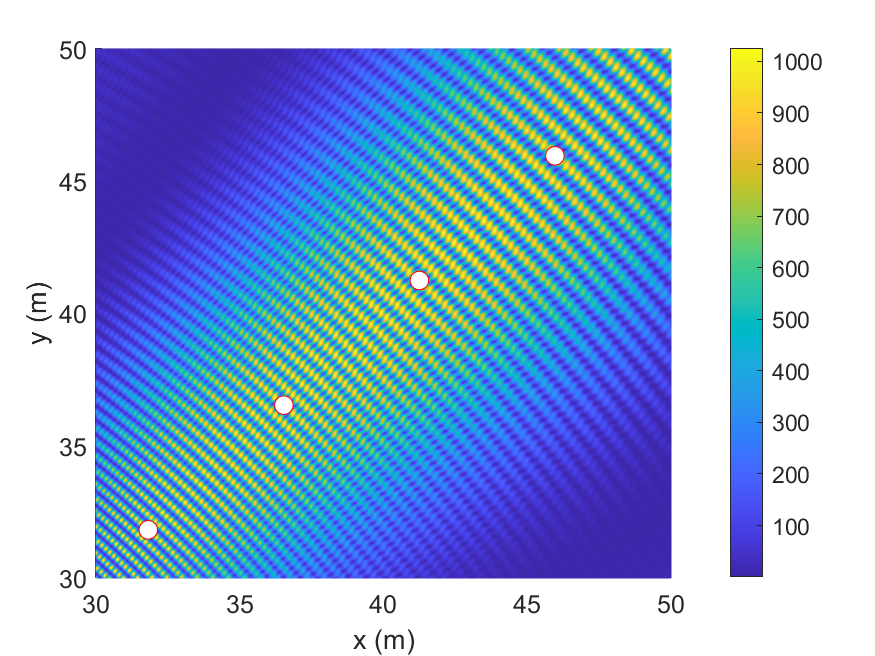}}
\caption{Beamforming gain for 3D users with similar angles.}
\label{fig:3D-pattern}
\end{minipage}\hfill
\begin{minipage}{0.5\textwidth}
\subfigure[Users at the same azimuth angle.] {
\label{fig_3D_SE:a}
\includegraphics[width=0.48\textwidth, height=0.45\textwidth]{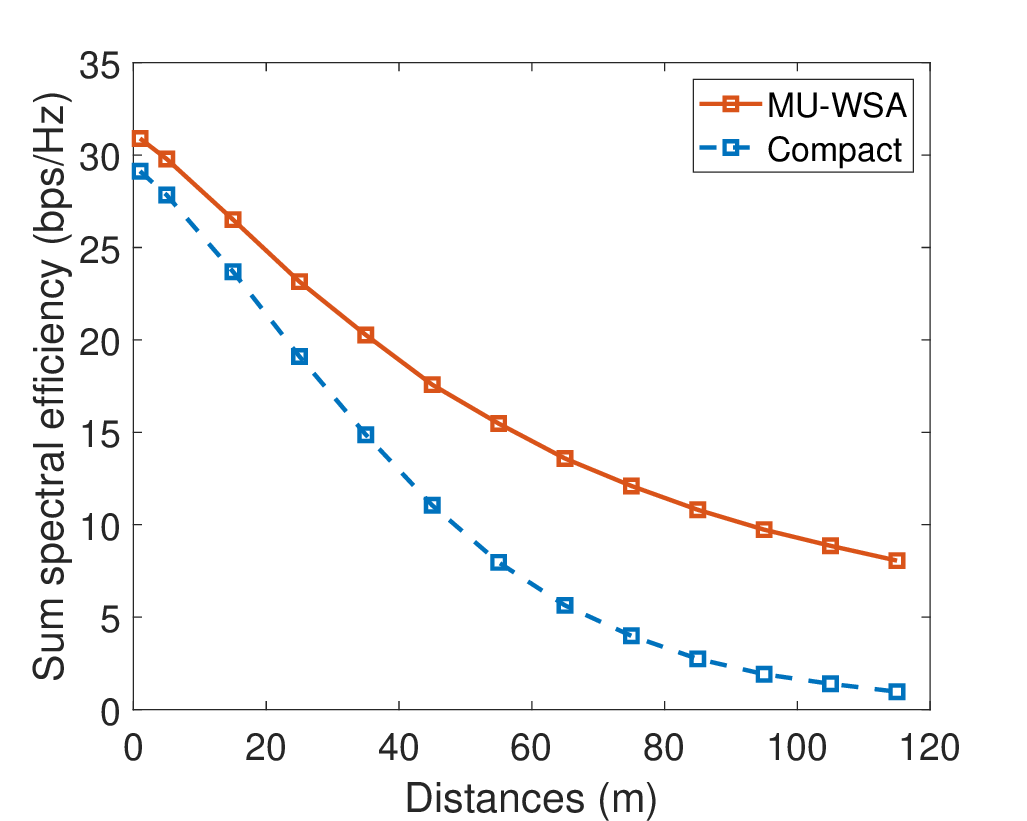}}\hfill
\subfigure[Users at different azimuth angles.]{
\label{fig_3D_SE:b}  
\includegraphics[width=0.48\textwidth, height=0.45\textwidth]{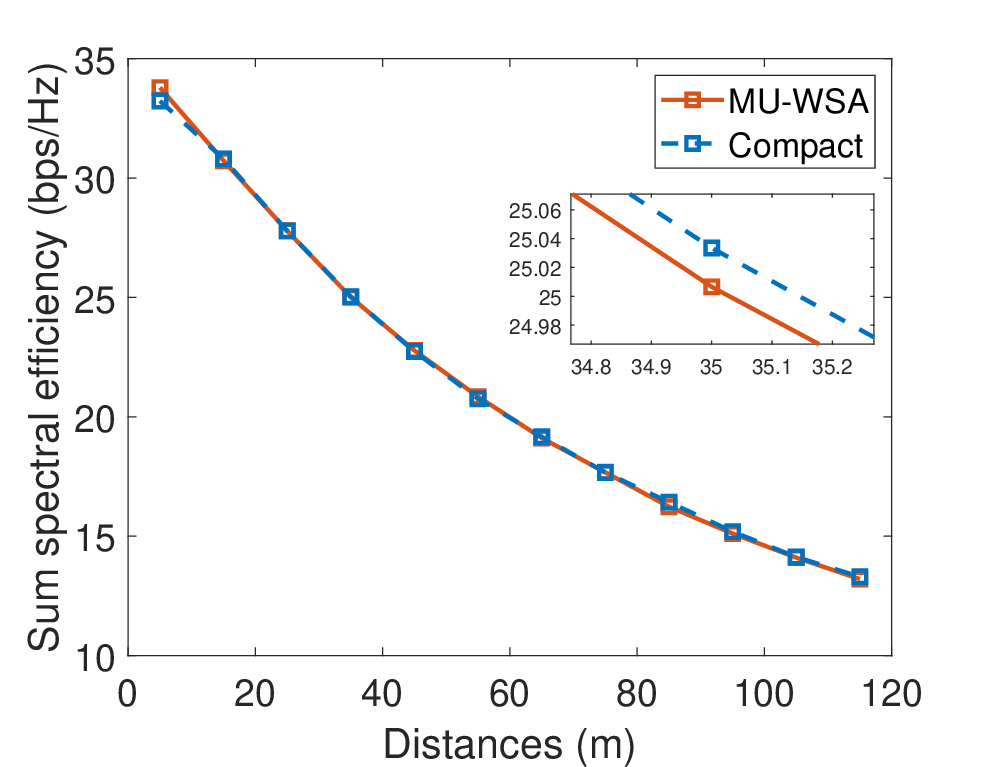}} 
\caption{3D scenario: optimal sum SE versus transmission distance.}
\label{fig:3D_SE}
\end{minipage}%
\end{figure*}

We first compare the performance of MU-WSA and the compact array in 2-D scenarios where the transmitters and the receivers are at approximately the same height such that only the azimuth angle dominates. In Fig. \ref{fig:2D-pattern}, we analyze the analog beampattern of MU-WSA and compact array where four users are located at the same azimuth angle, ranging from $1$ to $20$m, represented by the white dots shown in the figure. Comparing Fig. \ref{fig_2D:a} and \ref{fig_2D:b}, we observe that the far-field analog beam in the compact array fails to distinguish these users since the beamforming gain along the target direction is the same with no distance domain differentiation. While in the MU-WSA, due to the near-field focusing effect, the analog beam pattern depicts the feature of additional distance domain resolution, providing the opportunity to mitigate interference for users with similar angles. Moreover, as the beam is more focused toward the target position rather than a target angle, the power can be more concentrated. Additionally, we observe the existence of grating lobe from Fig.~\ref{fig_2D:b} which is caused by the widely-spaced subarrays. The grating lobe would introduce additional user interference into the system, which may require interference cancellation beamforming algorithms, such as the baseband BD algorithm depicted in Sec.~\ref{Sec_BB}.

In Fig.~\ref{fig:2D_SE}, we evaluate the performance of the sum SE versus the transmission distance. To focus on the performance of the antenna array architecture, we temporarily neglect the effect of different connection types and the analog beamforming algorithms, and only consider the fully-digital scenario where the upper bound of the sum SE is achieved by applying the digital BD algorithm~\cite{HBF,HBD}. Fig.~\ref{fig_2D_SE:a} shows the sum SE when users are at the same azimuth angle. We observe that with increasing distance between the users and the BS, the MU-WSA architecture first demonstrates a large performance improvement compared to the traditional compact architecture. By contrast, since the compact array fails to cancel the inter-user interference at the same angle, the sum SE drops to approximately zero because of the low SINR. When the communication distance increases to a certain point, the near-field beam-focusing effect vanishes, resulting in a smaller gap between the performance of the MU-WSA and the compact array. On the other hand, for users at different azimuth angles as shown in Fig.~\ref{fig_2D_SE:b}, both the MU-WSA and the compact array architecture are able to distinguish the users. However, due to the additional inter-user interference caused by the grating lobe in WSA, the SE in the compact array is slightly larger than that in the WSA array. Fortunately, due to the interference-cancellation-based BD algorithm, the difference is negligible across varying distances.

\subsubsection{3-D Cross Field Analysis}
In this section, we consider the 3D scenarios where both the azimuth angle $\phi$ and the elevation angle $\theta$ should be considered. Therefore, even when the users are at the same azimuth angle, the different elevation angles $\phi_u$ can support transmission to these users. However, when the distance between users and the BS becomes larger, the elevation angles of different users become approximately the same, i.e., $\phi_u \approx \phi, \forall u$. In this case, as shown in Fig.~\ref{fig:3D-pattern}, the beamforming gain of the compact and the MU-WSA show a similar behavior as the 2D case where the MU-WSA provides additional distance domain resolution and depicts the beam focusing pattern, at the cost of the grating lobe effect.

This effect can be further illustrated through the evaluation of the sum SE for users with the same azimuth angles. As shown in Fig.~\ref{fig:3D_SE}, we observe that when the transmission distance is small, the performance gap between the MU-WSA and the compact array architecture is relatively small due to the different elevation angles. However, as the communication distance increases, the MU-WSA starts to outperform the compact array. In Fig.~\ref{fig_3D_SE:b} where users are all at different azimuth angles, the sum SE is again similar for both architectures.

In Table~\ref{Table: comparison}, we conclude the MU-WSA and compact array architecture. First, due to the enlarged near-field, MU-WSA has the ability to distinguish users at the same angle while the compact array cannot. Secondly, the MU-WSA architecture takes advantage of the additional spatial degree of freedom such that it is capable of transmitting more data streams than the compact array. However, the MU-WSA will have a relatively large array aperture compared to the compact array and the channel complexity of the cross-field model is higher than the planar-wave model for the compact array architecture.

\begin{table}
\centering
\caption{Comparison of MU-WSA and compact array.}
\label{Table: comparison}%添加标题 设置标签
\begin{tabularx}{{\linewidth}}{|p{4cm}|X|p{1cm}|}
\hline
\centering\arraybackslash \textbf{Metric} & \centering\arraybackslash \textbf{WSA} & \centering\arraybackslash \textbf{Compact}\\
\hline
\makecell{ Channel model} & \makecell{CNFF or SWM} & \makecell{PWM}\\
\hline
\makecell{Array aperture} & \makecell{Relatively large} & \makecell{Small}\\
\hline
\makecell{ Ability to distinguish users \\ at the same angle} & \makecell{Yes} & \makecell{No}\\
\hline
\makecell{Capable of transmitting more \\ data streams than number\\ of multipath components} & \makecell{Yes} & \makecell{No}\\
\hline
\end{tabularx}
\end{table}

\subsection{Performance Analysis of Hybrid Beamforming Algorithms}
We then analyze the performance of the proposed algorithms in both MU-WSA and the compact array architecture.

\begin{figure} %H为当前位置，!htb为忽略美学标准，htbp为浮动图形
\centering %图片居中
\includegraphics[width = 0.45\textwidth,,height = 0.35\textwidth]{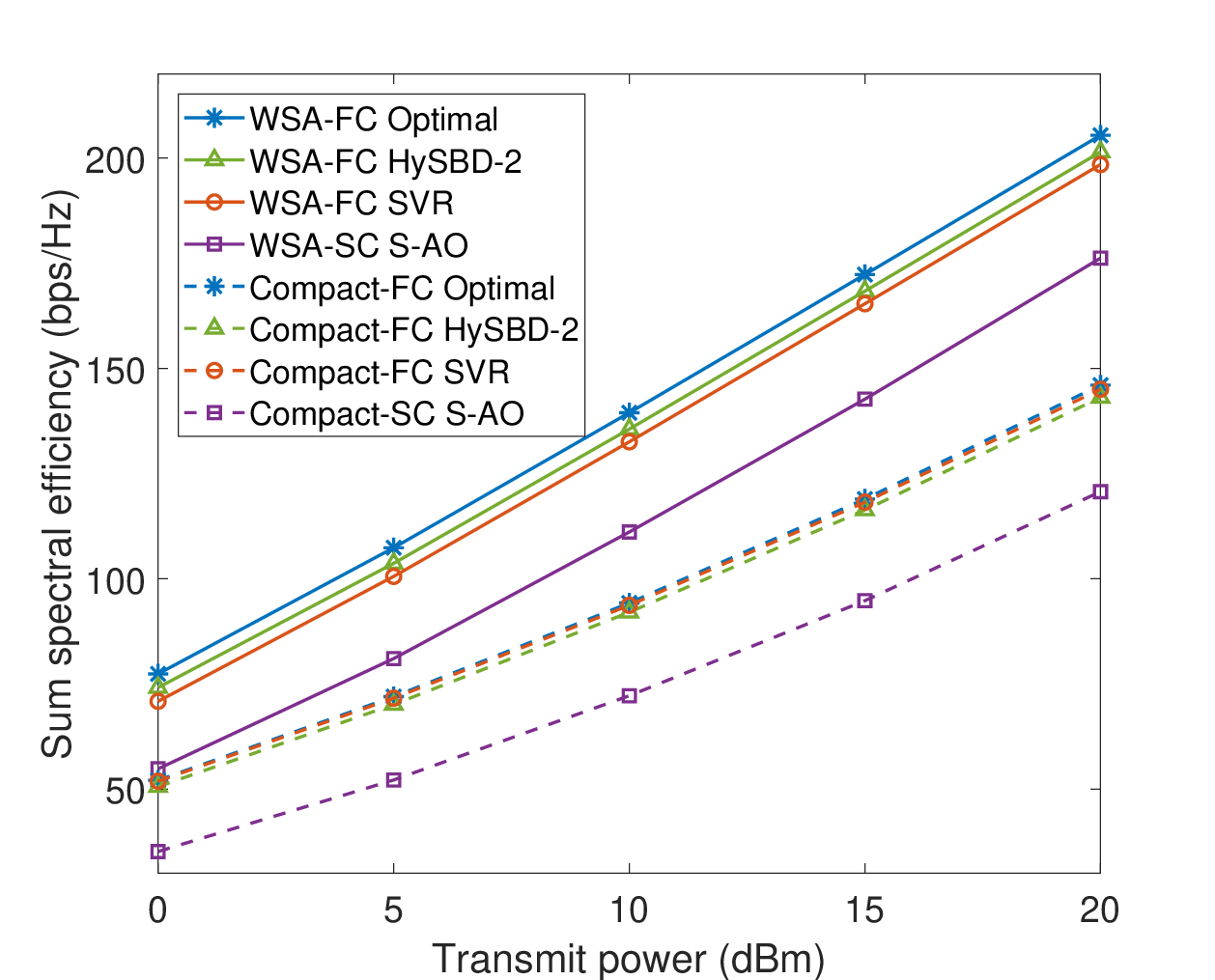} 
\caption{Sum spectral efficiency versus transmit power.} %最终文档中希望显示的图片标题
\label{fig:2D_power} %用于文内引用的标签
\end{figure}

In Fig.~\ref{fig:2D_power}, we first evaluate the sum SE of the proposed algorithms versus the transmit power. The users are distributed randomly within a sector of $120^{\circ}$ angle and a radius ranging from $1$ to $20$m. We compare the performance with the optimal upper bound and an SVD-based benchmark algorithm HySBD-2 proposed in~\cite{HBF}. From the figure, we observe that the MU-WSA architecture always outperforms the compact array regardless of the algorithms used. Specifically, comparing the upper bound of the two architectures, the MU-WSA improves the SE by over $60$\% at a power of $20$dBm. In the FC system, the proposed SVR algorithm can achieve over 95\% sum SE of the upper bound and the existing SVD-based algorithm, but with much lower computational complexity. Moreover, the proposed S-AO algorithm in SC achieves over 80\% of the sum SE in the FC system with a lower power consumption as the SC system reduces the number of phase shifters by $K^2$.

\begin{figure} %H为当前位置，!htb为忽略美学标准，htbp为浮动图形
\centering %图片居中
\includegraphics[width = 0.45\textwidth,height = 0.35\textwidth]{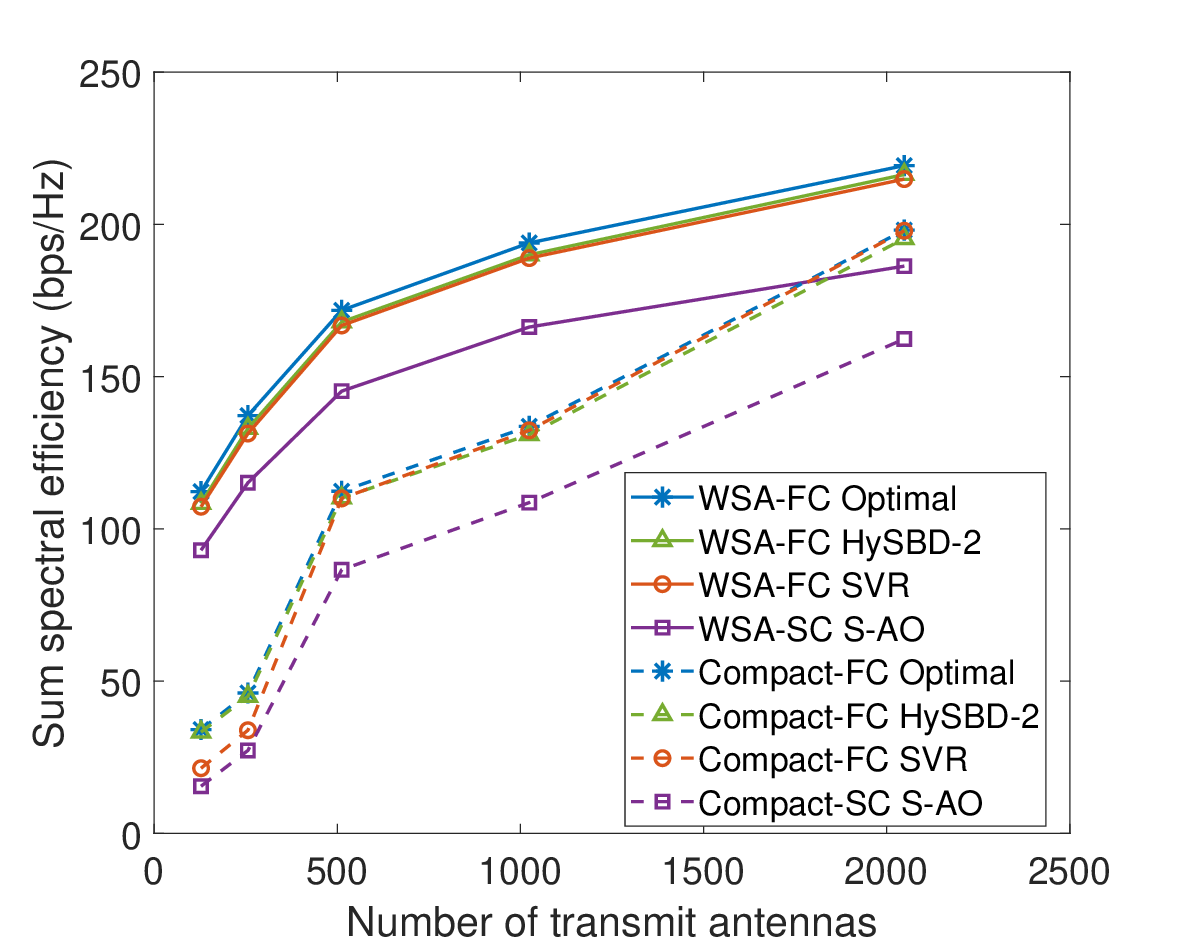} 
\caption{Sum spectral efficiency versus number of transmit antennas.} %最终文档中希望显示的图片标题
\label{fig:2D_antenna} %用于文内引用的标签
\end{figure}

Fig.~\ref{fig:2D_antenna} demonstrates the sum SE versus the number of transmit antennas. We observe that, at both FC and SC systems, when the number of transmit antennas is small, the performance gap between MU-WSA and the compact array is relatively large. As the number of antennas increases, the SE in MU-WSA improves at a more steady rate, while the SE in the compact array increases more rapidly with the growing number of antennas, resulting in a smaller gap of sum SE. This is because when the number of antennas is small, the beamwidth of the compact array is relatively large such that the power cannot be well focused towards a specific direction. When the number of antennas increases, a narrower beam can be generated, resulting in a much higher SE. In contrast, MU-WSA with its fixed array aperture maintains a relatively stable beamwidth and can conduct beam focusing to concentrate the power effectively even with fewer antennas, leading to a more consistent SE improvement. Therefore, the performance of the compact array is more dependent on the number of antennas compared to the MU-WSA.

\begin{figure} %H为当前位置，!htb为忽略美学标准，htbp为浮动图形
\centering %图片居中
\includegraphics[width = 0.45\textwidth,height = 0.38\textwidth]{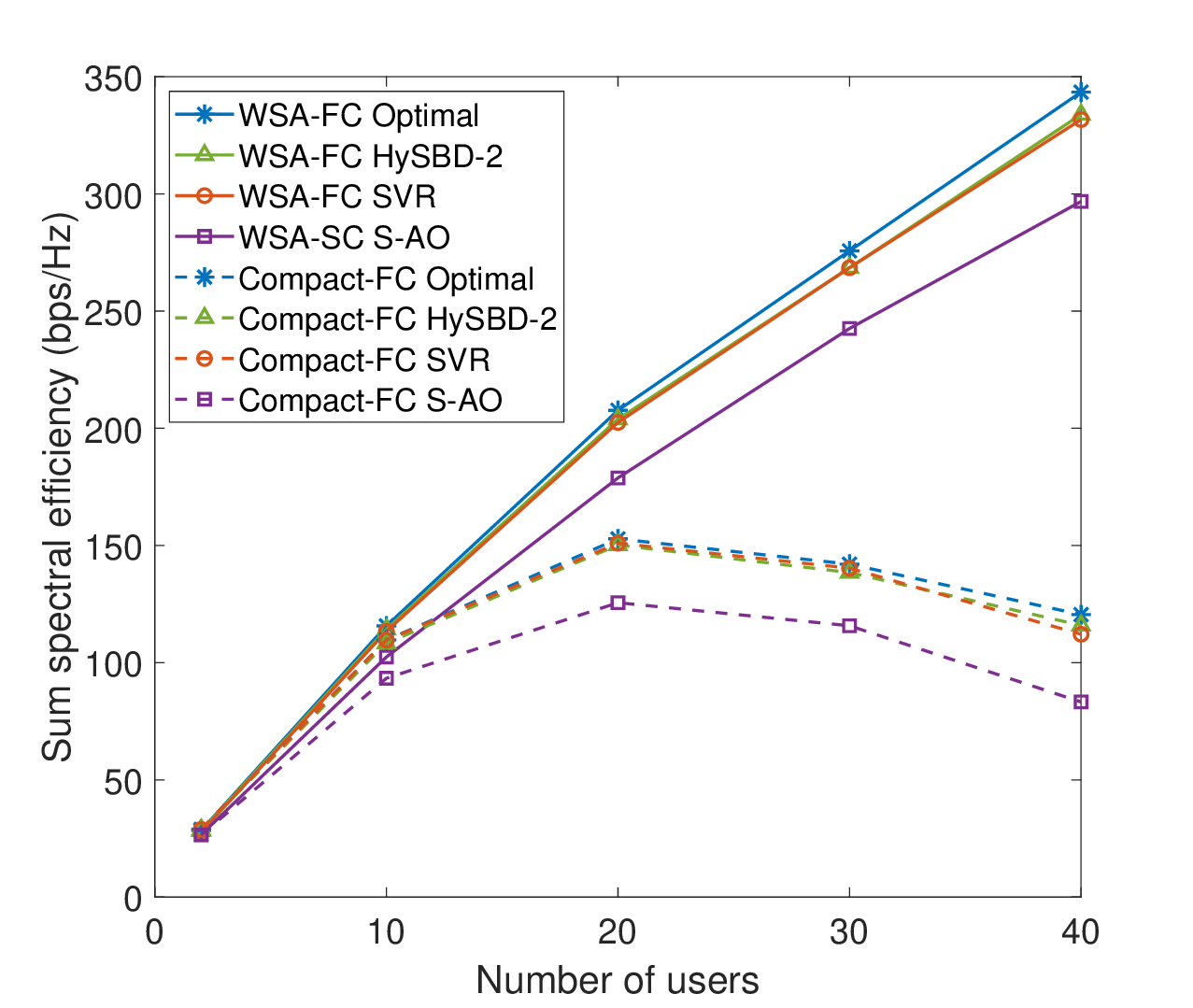} 
\caption{Sum spectral efficiency versus number of users.} %最终文档中希望显示的图片标题
\label{fig:2D_UEnum} %用于文内引用的标签
\end{figure}

In Fig.~\ref{fig:2D_UEnum}, we evaluate the sum SE versus the number of users. We observe that, when the number of users is small, i.e., the users are sparsely distributed in space, the performance of the MU-WSA and the compact array are similar. However, as the number of users increases, the compact array system experiences a rapid performance degradation. This is because, as the user density increases, users are more likely to be located at similar angles. Since the compact array relies on the angular resolution of far-field communication, it fails to mitigate interference from users at similar angles, thereby reducing the sum SE. In contrast, the MU-WSA system leverages the near-field distance domain, which enables efficient interference suppression even among users with identical angles. As a result, the WSA system can effectively serve all the users simultaneously, maintaining higher spectral efficiency.

In Table~\ref{Table: running time}, we analyze the computational complexity and the running time of the proposed algorithms. For SVR which utilizes a predefined steering vector, the precoder at each subarray for each user can be calculated through through simple calculation, resulting in the lowest complexity, i.e., $O(UK)$. While the benchmark algorithm HySBD-2~\cite{HBF} conducts SVD operation on the channel matrix of all users, resulting in a complexity of $O((UL_r)^2N_t)$. These two algorithms can be applied to the FC system while the proposed steering vector-based scheme has a much lower complexity. For the more complicated SC beamforming scheme, the proposed S-AO algorithm is of the highest complexity since it requires the calculation of the inverse of the matrix $\mathbf{B}_j \in \mathbb{C}^{\frac{(K-1)L_t}{K}\times \frac{(K-1)L_t}{K}}$ in (\ref{eq: Bj}), which has a complexity of $O((K-1)^3L_t^3/K^3)$, and the SVD operation for $\mathbf{X}_j \in \mathbb{C}^{\frac{N_t}{K} \times \frac{N_t}{K}}$, which has a complexity of $O((\frac{N_t}{K})^3)$. Therefore, the overall complexity after $K$ iterations is $O(((K-1)^3L_t^3+N_t^3)/K^2)$. The running time of the algorithms with MatLab R2022a is also summarized to directly demonstrate the complexity. We observe that the running time corresponds to the complexity analysis, where SVR has the lowest complexity. While S-AO has the highest complexity but is able to solve the more complex SC problem with acceptable performance loss.

\section{Conclusion}
\label{section_conclusion}

\begin{table}
\centering
\caption{Algorithm complexity and running time in MATLAB R2022a.
}
\label{Table: running time}%添加标题 设置标签
\begin{tabularx}{\linewidth}{|>{\centering\arraybackslash}p{2.3cm}|>{\centering\arraybackslash}X|>{\centering\arraybackslash}p{1.75cm}|}
\hline
\textbf{Algorithm}& \textbf{Complexity for analog stage} &\textbf{Running time}\\
\hline
SVR (FC) & $O(UK)$ & around 0.02s\\
\hline
HySBD-2 (FC)~\cite{HBF}& $O(U^2L_r^2N_t)$ & around 0.18s\\
\hline
S-AO (SC) & $O(((K-1)^3L_t^3+N_t^3)/K^2)$ & around 0.35s\\
\hline
\end{tabularx}
\end{table}
In this paper, we extended the WSA hybrid beamforming architecture into the multi-user scenarios, which can utilize both the angular and distance domain resolutions. Since PWM cannot be applied in the enlarged near-field region of MU-WSA, we investigated the CNFF channel model. Then, by considering both SC and FC systems, we formulated a joint optimization problem to maximize the sum SE. We then decomposed the problem into the design of the MU-WSA architecture, i.e., the subarray spacing $d_s$ and the number of subarrays $K$, and the hybrid beamformers. Firstly, based on the analysis of the $(K.d_s)$ pair, we derive the optimal subarray spacing for the architecture design. Then, for SC systems where the analog precoder has a block-diagonal format, we propose the S-AO algorithm to iteratively optimize each submatrix in the analog precoder. In the FC system, a low-complexity SVR algorithm is proposed where the analog beamformers are found through the reconstruction of the far-field steering vectors. We evaluate the beam pattern of MU-WSA and the compact array in 2D and 3D cases. We show that in 2D cases, the MU-WSA outperforms the traditional compact array when users are located at the same azimuth angle. While in 3D scenarios, the compact array and the MU-WSA exhibit comparable performance at short distances, but as the distance increases, MU-WSA depicts a gradual superiority over the compact array. When users are located at different angles, the two architectures can achieve similar performance. Therefore, to conclude, the MU-WSA architecture is preferred when the users are more likely to locate at similar angles or directions, while both the MU-WSA and the compact array are feasible when users tend to be scattered in the angle domain. Furthermore, we compare the performance of the proposed S-AO and SVR algorithms. We show that the proposed S-AO algorithm in the SC system can achieve over 80\% of the sum SE in the FC system but with lower hardware complexity. Moreover, the proposed SVR achieves over 95\% of the upper bound of SE, but much lower computational complexity.

\section*{Appendix}
To prove that $g'(d_s) < 0$ when $D^{um} \geq 2\sqrt{2}(\sqrt{N_r}-1) S_t^{\text{max}}, \forall u,m$, we first show $|\beta_{k,i,j}-\beta_{m,i,j}|\leq \pi$. Based on equation (\ref{Eq:distance_tay}) and (\ref{Eq:WSA angle}), we have
\begin{equation}
    \sin \theta_k \cos \phi_k = \frac{x_u-x_k}{D^{uk}}.
\end{equation} 
Therefore, $\beta_{k,i,j}$ can be represented as
\begin{equation}
    \beta_{k,i,j} = \pi\left[ (j_1-i_1)\frac{x_u-x_K}{D^{uk}}+(j_2-i_2)\frac{z_u-z_k}{D^{uk}}\right],
\end{equation}
where $(i_1,j_1)$ and $(i_2,j_2)$ are the x- and z-coordinate of antenna $i$ and $j$, respectively. 
We denote $h(d) = \frac{1}{\pi}|\beta_{k,i,j}-\beta_{m,i,j}|$ where $d=\frac{(n-1)}{2}\lambda+d_s$. Next, we prove $h(d)<1$. Specifically, we can derive
\begin{subequations}
\begin{align}
     & h(d) = |\beta_{k,i,j}-\beta_{m,i,j})|/\pi \notag \\
     &=|(j_1-i_1)[(x_u-x_k)(\frac{1}{D^{uk}}-\frac{1}{D^{um}})+(m_x-k_x)\frac{d}{D^{um}}] \notag \\
     & +(j_2-i_2)[(z_u-z_k)(\frac{1}{D^{uk}}-\frac{1}{D^{um}})+(m_z-k_z)\frac{d}{D^{um}}]| \notag \\
     & \leq |(j_1-i_1)[(x_u-x_k)(\frac{1}{D^{uk}}-\frac{1}{D^{um}})+(m_x-k_x)\frac{d}{D^{um}}]|\notag \\
     & +|(j_2-i_2)[(z_u-z_k)(\frac{1}{D^{uk}}-\frac{1}{D^{um}})+(m_z-k_z)\frac{d}{D^{um}}]|\notag \\
     & \leq (\sqrt{Nr}-1)[\left|x_u-x_k\right|\cdot\left|\frac{1}{D^{uk}}-\frac{1}{D^{um}}\right|+(\sqrt{K}-1)\frac{d}{D^{um}}] \notag \\
     & + (\sqrt{Nr}-1)[\left|z_u-z_k\right|\cdot \left|\frac{1}{D^{uk}}-\frac{1}{D^{um}}\right|+(\sqrt{K}-1)\frac{d}{D^{um}}] \notag \\
     &=(\sqrt{Nr}-1)\left[ \left(|x_u-x_k|+|z_u-z_k|\right)\cdot\left|\frac{1}{D^{uk}}-\frac{1}{D^{um}}\right|\right.\notag \\
     & \quad \quad \quad \quad \quad \quad \left. +2(\sqrt{K}-1)\frac{d}{D^{um}}\right] \tag{43}
\end{align}
\end{subequations}
Since $|D^{uk}-D^{um}|\leq S_t$, then by letting $D^{um} = D^{uk}+S_t$, we have
\begin{subequations}
    \begin{align}
    \begin{split}
        h(d) & \leq (\sqrt{Nr}-1)\left[\frac{(\left|x_u-x_k\right|+\left|z_u-z_k\right|)}{D^{uk}}\frac{S_t}{D^{uk}+S_t} \right.\\
        & \quad \quad \quad \quad \quad \quad \left. +2(\sqrt{K}-1)\frac{d}{D^{um}}\right]
    \end{split}\\
        & \leq (\sqrt{Nr}-1)(\frac{\sqrt{2}S_t}{D^{um}}+2(\sqrt{K}-1)\frac{d}{D^{um}}),\label{eq:inequality}
    \end{align}
\end{subequations}
where (\ref{eq:inequality}) is obtained from the Cauchy–Schwarz inequality that
\begin{equation}
    |a|+|b|\leq\sqrt{2}\sqrt{a^2+b^2}\leq \sqrt{a^2+b^2+c^2}.
\end{equation}
Now, to prove $g(d_s)<1$, we need to equivalently prove
\begin{equation}
    \sqrt{2}S_t+2(\sqrt{K}-1)d_s \leq \frac{D^{um}}{\sqrt{Nr}-1}.
\end{equation}
Since $\sqrt{2}(\sqrt{K}-1)d_s < S_t$, we have 
\begin{equation}
    h(d_s) \leq 2\sqrt{2} S_t.
\end{equation}
Therefore, when the distance between the subarray and the user satisfies $D^{um} \geq 2\sqrt{2}(\sqrt{N_r}-1) S_t, \forall u,m$, the result of $|\beta_{k,i,j}-\beta_{m,i,j}|\leq \pi$ is ensured, i.e., $\sin{(\beta_{k,i,j}-\beta_{m,i,j})}$ has the same sign with $\beta_{k,i,j}-\beta_{m,i,j}$. e denote $p(d) = (\beta_{k,i,j}-\beta_{m,i,j})/\pi$, then the original problem is transferred to proving that
\begin{equation}
\label{eq:delta}
    \Delta(d) = 2\sum_{k<m}[-p(d)\cdot (\frac{\partial p(d)}{\partial d})] < 0.
\end{equation}
Through calculation, we can derive
\begin{equation}
\begin{aligned}
    & \frac{\partial p(d)}{\partial d} = -\frac{\left(j_{2} - i_{2}\right) \left(1 - m_{z}\right) + \left(j_{1} - i_{1}\right) \left(1 - m_{x}\right)}{D^{um}} \\
     & + \frac{\left(j_{2} - i_{2}\right) \left(1 - k_{z}\right) + \left(j_{1} - i_{1}\right) \left(1 - k_{x}\right)}{D^{uk}}\\
    & \resizebox{1\hsize}{!}{$+ \frac{\left[\left(j_{2} - i_{2}\right) \left(z_{u} - z_m\right) + \left(j_{1} - i_{1}\right) \left(x_{u} - x_m\right)\right]\left[ \left(m_{z} - 1\right) \left(z_m-z_u\right) +  \left(m_{x} - 1\right) \left(x_m - x_{u}\right)\right]}{(D^{um})^3}$}\\
    & \resizebox{1\hsize}{!}{$- \frac{\left[\left(j_{2} - i_{2}\right) \left(z_{u} - z_k\right) + \left(j_{1} - i_{1}\right) \left(x_{u} - x_k\right)\right]\left[ \left(k_{z} - 1\right) \left(z_k-z_u\right) +  \left(k_{x} - 1\right) \left(x_k - x_{u}\right)\right]}{(D^{uk})^3}$}.
\end{aligned}
\end{equation}
Without loss of generality, we assume $p(d)>0$. Equivalently, we have
\begin{equation}
\begin{aligned}
    & \frac{(j_1-i_1)(x_u-x_k)+(j_2-i_2)(z_u-z_k)}{D^{uk}}\\
    & -\frac{(j_1-i_1)(x_u-x_m)+(j_2-i_2)(z_u-z_m)}{D^{um}}>0\\
    & \Leftrightarrow  \frac{(j_1-i_1)[(k_x-1)d-x_u]+(j_2-i_2)[(k_z-1)d-z_u']}{D^{uk}}\\
    &<\frac{(j_1-i_1)[(m_x-1)d-x_u]+(j_2-i_2)[(m_z-1)d-z_u']}{D^{um}}\\
    &\Leftrightarrow  [\frac{\left(j_{1} - i_{1}\right) \left(k_{x}-1\right) + \left(j_{2} - i_{2}\right) \left(k_{z}-1\right)}{D^{uk}}\\
    & -\frac{\left(j_{1} - i_{1}\right) \left(m_{x}-1\right) + \left(j_{2} - i_{2}\right) \left(m_{x}-1\right)}{D^{um}}]d\\
    & <\frac{(j_1-i_1)x_u}{D^{uk}}+\frac{(j_2-i_2)z_u'}{D^{uk}}-\frac{(j_1-i_1)x_u}{D^{um}}-\frac{(j_2-i_2)z_u'}{D^{um}},
\end{aligned}
\end{equation}
where $z_u' = z_u-h_t$. Therefore, we can obtain
\begin{subequations}
\begin{align}
     & {\frac{\partial p(d)}{\partial d} > \left[\left(j_{2} - i_{2}\right) \left(z_{u}' - z_m\right) + \left(j_{1} - i_{1}\right) \left(x_{u} - x_m\right)\right]} \notag \\
     & \times \left[ \left(m_{z} - 1\right) \left(z_m-z_u'\right) +  \left(m_{x} - 1\right) \left(x_m - x_{u}\right)\right]/{(D^{um})^3} \notag \\
     & - \left[\left(j_{2} - i_{2}\right) \left(z_{u}' - z_k\right) + \left(j_{1} - i_{1}\right) \left(x_{u} - x_k\right)\right]\notag\\
     & \times \left[ \left(k_{z} - 1\right) \left(z_k-z_u'\right) +  \left(k_{x} - 1\right) \left(x_k - x_{u}\right)\right]/{(D^{uk})^3} \notag \\
     & \resizebox{1\hsize}{!}{$-\frac{1}{d}[(j_1-i_1)x_u(\frac{1}{D^{uk}}-\frac{1}{D^{um}})+(j_2-i_2)z_u'(\frac{1}{D^{uk}}-\frac{1}{D^{um}})]$}\notag\\
     & \resizebox{1\hsize}{!}{$= [\frac{1}{(D^{uk})^3}-\frac{1}{(D^{um})^3}][(j_1-i_1)(x_u-x_k)+(j_2-i_2)(z_u'-z_k)]$}\notag \\
     & \times [(k_x-m_x)(x_u-x_k)+(k_z-m_z)(z_u'-z_k)]\notag\\
     & + \frac{[(j_1-i_1)(x_u-x_k)+(j_2-i_2)(z_u'-z_k)]}{(D^{um})^3}\notag\\
     & \times [(m_x-1)(m_x-k_x)+(m_z-1)(m_z-k_z)]d \notag \\
     & -\frac{d}{(D^{um})^3}[(j_1-i_1)(m_x-k_x)+(j_2-i_2)(m_z-k_z)] \notag\\
     &\times [(m_z-1)(z_k-z_u')+(m_x-1)(x_k-x_u)\notag\\
     & +(m_z-1)(m_z-k_z)d+(m_x-1)(m_x-k_x)d].\tag{51}
\end{align}
\end{subequations}
Since $p(d)>0$, we have
\begin{subequations}
\begin{align}
    &(j_1-i_1)[(x_u-x_k)(\frac{1}{D^{uk}}-\frac{1}{D^{um}})+(m_x-k_x)\frac{d}{D^{um}}] \notag\\
     &\resizebox{0.9\hsize}{!}{$+(j_2-i_2)[(z_u'-z_k)(\frac{1}{D^{uk}}-\frac{1}{D^{um}})+(m_z-k_z)\frac{d}{D^{um}}]>0$} \notag\\
     &\resizebox{0.9\hsize}{!}{$\Leftrightarrow  (\frac{1}{D^{uk}}-\frac{1}{D^{um}})[(j_1-i_1)(x_u-x_k)+(j_2-i_2)(z_u'-z_k)]$}\notag\\
     & > \frac{d}{D^{um}}[(j_1-i_1)(k_x-m_x)+(j_2-i_2)(k_z-m_z)]. \tag{52}
\end{align}
\end{subequations}
Therefore, we can further derive
\begin{subequations}
\begin{align}
        & \frac{\partial p(d)}{\partial d} > [(j_1-i_1)(k_x-m_x)+(j_2-i_2)(k_z-m_z)]\notag \\
        & \times [(k_x-m_x)(x_u-x_k)+(k_z-m_z)(z_u'-z_k)] \notag \\
        & \times [(\frac{1}{D^{uk}})^2+(\frac{1}{D^{um}})^2+\frac{1}{D^{uk}D^{um}}]\frac{d}{D^{um}}\notag \\
        & - \frac{d}{(D^{um})^3}[(j_1-i_1)(x_u-x_k)+(j_2-i_2)(z_u'-z_k)] \notag\\
        & \times [(m_x-1)(m_x-k_x)+(m_z-1)(m_z-k_z)]\notag \\
     & +[(m_z-1)(z_u'-z_k)+(m_x-1)(x_u-x_k)\notag \\
     & +(m_z-1)(m_z-k_z)d+(m_x-1)(m_x-k_x)d]\notag \\
     & \times \frac{[(j_1-i_1)(m_x-k_x)+(j_2-i_2)(m_z-k_z)]d}{(D^{um})^3}\notag \\
        & \resizebox{1\hsize}{!}{$= \Big\{(j_1-i_1)[(x_u-x_k)(k_x-m_x)^2+(m_x-k_x)(m_z-k_z)(z_u'-z_k)]$} \notag \\
        &\resizebox{1\hsize}{!}{$+(j_2-i_2)[(z_u'-z_k)(k_z-m_z)^2+(m_z-k_z)(m_x-k_x)(x_u-x_k)]\Big\}$} \notag \\
        & \times [(\frac{1}{D^{uk}})^2+\frac{1}{D^{uk}D^{um}}]\frac{d}{D^{um}} \notag\\
        & +(j_1-i_1)\Big[(x_u-x_k)\frac{(k_x-m_x)^2d-(m_z-1)(m_z-k_z)d}{(D^{um})^3}\notag \\
        & +(m_x-k_x)\frac{(m_z-k_z)(z_u'-z_k)d+(m_z-1)(z_u'-z_k)d}{(D^{um})^3} \notag \\
        & \resizebox{0.94\hsize}{!}{$+(m_x-k_x)\frac{(m_z-1)(m_z-k_z)d^2+(m_x-k_x)(m_x-1)d^2}{(D^{um})^3}\Big]$} \notag \\
        & +(j_2-i_2)\Big[(z_u'-z_k)\frac{(k_z-m_z)^2d-(m_x-1)(m_x-k_x)d}{(D^{um})^3} \notag \\
        & +(m_z-k_z) \frac{(m_x-k_x)(x_u-x_k)d+(m_x-1)(x_u-x_k)d}{(D^{um})^3} \notag \\
        & \resizebox{0.95\hsize}{!}{$+(m_z-k_z)\frac{(m_z-1)(m_z-k_z)d^2+(m_x-1)(m_x-k_x)d^2}{(D^{um})^3}\Big].$} \tag{53}
\end{align}
\end{subequations}
Strict proof of the positivity of eq.~(53) requires complex case studies to analyze the signs of $i,j,k$ and $m$, as well as the location of the user. However, since the function is of high complexity, a formal proof is challenging. Nevertheless, we provide some insights here which suggests the formula to be positive.
Since $k<m$ as stated in eq.~(\ref{eq:delta}), we generally have $(m_x-k_x)(m_z-k_z)>0$. Therefore, we can obtain 
\begin{equation}
\begin{aligned}
    & \resizebox{1\hsize}{!}{$\Big\{(j_1-i_1)[(x_u-x_k)(k_x-m_x)^2+(m_x-k_x)(m_z-k_z)(z_u'-z_k)]$}\\
    & \resizebox{1\hsize}{!}{$+(j_2-i_2)[(z_u'-z_k)(k_z-m_z)^2+(m_z-k_z)(m_x-k_x)(x_u-x_k)\Big\}$}\\
    & \times [(\frac{1}{D^{uk}})^2+\frac{1}{D^{uk}D^{um}}]\frac{d}{D^{um}}\\
    & > p(d)[(\frac{1}{D^{uk}})^2+\frac{1}{D^{uk}D^{um}}]\frac{d}{D^{um}} >0.
\end{aligned}
\end{equation}%
Moreover, since we have 
\begin{equation}
\begin{aligned}
    & D^{um}>2\sqrt{2}(\sqrt{Nr}-1)S_t>4(\sqrt{Nr}-1)(\sqrt{K}-1)d,\\
    & \text{max}\{|j_1-i_1|,|j_2-i_2|\}<\sqrt{N_r}-1,\\
    & \text{max}\{|k_x-m_x|,|k_z-m_z|\}<\sqrt{K}-1,
\end{aligned}
\end{equation}
we can then approximately obtain
\begin{equation}
\begin{aligned}
    &\resizebox{1\hsize}{!}{$\frac{\partial p(d)}{\partial (d)} > \{(j_1-i_1)[(x_u-x_k)(\frac{1}{D^{uk}}-\frac{1}{D^{um}})+(m_x-k_x)\frac{d}{D^{um}}]$}\\
     & +(j_2-i_2)[(z_u'-z_k)(\frac{1}{D^{uk}}-\frac{1}{D^{um}})+(m_z-k_z)\frac{d}{D^{um}}]\}\\
     & \times [(\frac{1}{D^{uk}})^2+\frac{1}{D^{uk}D^{um}}+\frac{1}{D^{um}}]\\
    & = [(\frac{1}{D^{uk}})^2+\frac{1}{D^{uk}D^{um}}+\frac{1}{D^{um}}]p(d)>0.
\end{aligned}
\end{equation}
Finally, we obtain $\Delta(d) = 2\sum_{k<m}[-p(d)\cdot (\frac{\partial p(d)}{\partial d})] < 0$. $\hfill \blacksquare$

\begin{figure}[htbp]
    \centering
    {\includegraphics[width = 0.5\textwidth]{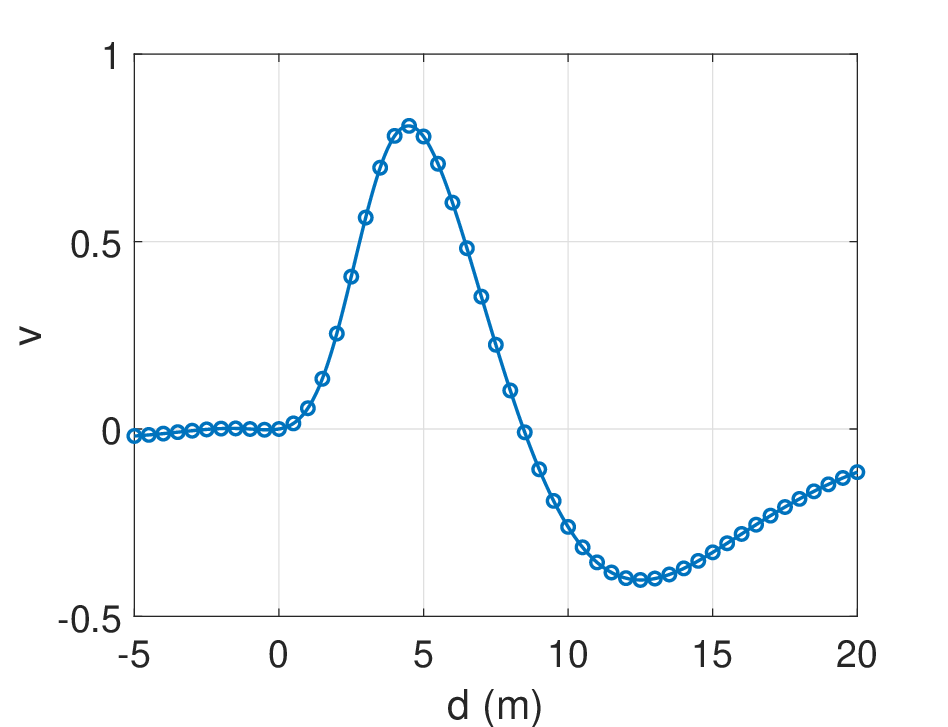}}
    \caption{$V(d)$ versus $d$.}
    \label{Fig.10}
\end{figure}

To further illustrate the proof, we show the numerical results of $V(d) = p(d)\frac{\partial p(d)}{p(d)}$ which can be obtained using mathematical software such as MATLAB with randomly chosen $i,j,k$, and $m$. We fix the transmission distance at $15$~m, and plot $V(d)$ in Fig.~\ref{Fig.10}. We observe that when $d$ increases from $0$, the function $V(d)$ is initially positive, indicating an increasing capacity. While when $d$ increases further such that the array aperture becomes comparable to the transmission distance, function $V$ gradually decreases and becomes negative, indicating a decline in capacity.

\bibliographystyle{IEEEtran}
\bibliography{reference}

\begin{thebibliography}{10}
\providecommand{\url}[1]{#1}
\csname url@samestyle\endcsname
\providecommand{\newblock}{\relax}
\providecommand{\bibinfo}[2]{#2}
\providecommand{\BIBentrySTDinterwordspacing}{\spaceskip=0pt\relax}
\providecommand{\BIBentryALTinterwordstretchfactor}{4}
\providecommand{\BIBentryALTinterwordspacing}{\spaceskip=\fontdimen2\font plus
\BIBentryALTinterwordstretchfactor\fontdimen3\font minus \fontdimen4\font\relax}
\providecommand{\BIBforeignlanguage}[2]{{%
\expandafter\ifx\csname l@#1\endcsname\relax
\typeout{** WARNING: IEEEtran.bst: No hyphenation pattern has been}%
\typeout{** loaded for the language `#1'. Using the pattern for}%
\typeout{** the default language instead.}%
\else
\language=\csname l@#1\endcsname
\fi
#2}}
\providecommand{\BIBdecl}{\relax}
\BIBdecl

\bibitem{overview}
I.~F. Akyildiz, C.~Han, Z.~Hu, S.~Nie, and J.~M. Jornet, ``Terahertz band communication: An old problem revisited and research directions for the next decade,'' \emph{IEEE Transactions on Communications}, vol.~70, no.~6, pp. 4250--4285, 2022.

\bibitem{combat}
I.~F. Akyildiz, C.~Han, and S.~Nie, ``Combating the distance problem in the millimeter wave and terahertz frequency bands,'' \emph{IEEE Communications Magazine}, vol.~56, no.~6, pp. 102--108, 2018.

\bibitem{UM-MIMO}
I.~F. Akyildiz and J.~M. Jornet, ``Realizing ultra-massive {MIMO} (1024×1024) communication in the (0.06–10) terahertz band,'' \emph{Nano Communication Networks}, vol.~8, pp. 46--54, 2016.

\bibitem{Challenges}
C.~Han, L.~Yan, and J.~Yuan, ``Hybrid beamforming for terahertz wireless communications: Challenges, architectures, and open problems,'' \emph{IEEE Wireless Communications}, vol.~28, no.~4, pp. 198--204, 2021.

\bibitem{Survey}
S.~A. Busari, K.~M.~S. Huq, S.~Mumtaz, L.~Dai, and J.~Rodriguez, ``Millimeter-wave massive {MIMO} communication for future wireless systems: A survey,'' \emph{IEEE Communications Surveys \& Tutorials}, vol.~20, no.~2, pp. 836--869, 2018.

\bibitem{modeling}
W.~Tang, M.~Z. Chen, X.~Chen, J.~Y. Dai, Y.~Han, M.~Di~Renzo, Y.~Zeng, S.~Jin, Q.~Cheng, and T.~J. Cui, ``Wireless communications with reconfigurable intelligent surface: Path loss modeling and experimental measurement,'' \emph{IEEE Transactions on Wireless Communications}, vol.~20, no.~1, pp. 421--439, 2021.

\bibitem{Spacespartial}
O.~E. Ayach, S.~Rajagopal, S.~Abu-Surra, Z.~Pi, and R.~W. Heath, ``Spatially sparse precoding in millimeter wave {MIMO} systems,'' \emph{IEEE Transactions on Wireless Communications}, vol.~13, no.~3, pp. 1499--1513, 2014.

\bibitem{HAD}
F.~Sohrabi and W.~Yu, ``Hybrid analog and digital beamforming for mmwave ofdm large-scale antenna arrays,'' \emph{IEEE Journal on Selected Areas in Communications}, vol.~35, no.~7, pp. 1432--1443, 2017.

\bibitem{HBD}
W.~Ni and X.~Dong, ``Hybrid block diagonalization for massive multiuser {MIMO} systems,'' \emph{IEEE Transactions on Communications}, vol.~64, no.~1, pp. 201--211, 2016.

\bibitem{HBF}
X.~Wu, D.~Liu, and F.~Yin, ``Hybrid beamforming for multi-user massive {MIMO} systems,'' \emph{IEEE Transactions on Communications}, vol.~66, no.~9, pp. 3879--3891, 2018.

\bibitem{subconnected}
Z.~Zhang, X.~Wu, and D.~Liu, ``Joint precoding and combining design for hybrid beamforming systems with subconnected structure,'' \emph{IEEE Systems Journal}, vol.~14, no.~1, pp. 184--195, 2020.

\bibitem{Dual}
Y.~Zhang, J.~Du, Y.~Chen, X.~Li, K.~M. Rabie, and R.~Khkrel, ``Dual-iterative hybrid beamforming design for millimeter-wave massive multi-user {MIMO} systems with sub-connected structure,'' \emph{IEEE Transactions on Vehicular Technology}, vol.~69, no.~11, pp. 13\,482--13\,496, 2020.

\bibitem{energy_sub}
X.~Gao, L.~Dai, S.~Han, C.-L. I, and R.~W. Heath, ``Energy-efficient hybrid analog and digital precoding for mmwave {MIMO} systems with large antenna arrays,'' \emph{IEEE Journal on Selected Areas in Communications}, vol.~34, no.~4, pp. 998--1009, 2016.

\bibitem{subconnected1}
Y.~Hu, H.~Qian, K.~Kang, X.~Luo, and H.~Zhu, ``Joint precoding design for sub-connected hybrid beamforming system,'' \emph{IEEE Transactions on Wireless Communications}, vol.~23, no.~2, pp. 1199--1212, 2024.

\bibitem{subconnected2}
J.~Du, Z.~Wang, Y.~Zhang, Y.~Guan, and L.~Jin, ``Multi-user hybrid precoding for mmwave massive {MIMO} systems with sub-connected structure,'' \emph{EURASIP Journal on Wireless Communications and Networking}, vol. 2021, no.~1, p. 157, Jul 2021.

\bibitem{InterIntra}
L.~Yan, Y.~Chen, C.~Han, and J.~Yuan, ``Joint inter-path and intra-path multiplexing for terahertz widely-spaced multi-subarray hybrid beamforming systems,'' \emph{IEEE Transactions on Communications}, vol.~70, no.~2, pp. 1391--1406, 2022.

\bibitem{Multi-Ray}
C.~Han, A.~O. Bicen, and I.~F. Akyildiz, ``Multi-ray channel modeling and wideband characterization for wireless communications in the terahertz band,'' \emph{IEEE Transactions on Wireless Communications}, vol.~14, no.~5, pp. 2402--2412, 2015.

\bibitem{measurement}
K.~Guan, B.~Peng, D.~He, J.~M. Eckhardt, S.~Rey, B.~Ai, Z.~Zhong, and T.~Kürner, ``Measurement, simulation, and characterization of train-to-infrastructure inside-station channel at the terahertz band,'' \emph{IEEE Transactions on Terahertz Science and Technology}, vol.~9, no.~3, pp. 291--306, 2019.

\bibitem{multifocus}
R.~Liu and K.~Wu, ``Antenna array for amplitude and phase specified near-field multifocus,'' \emph{IEEE Transactions on Antennas and Propagation}, vol.~67, no.~5, pp. 3140--3150, 2019.

\bibitem{different}
Y.~{Liu}, J.~{Xu}, Z.~{Wang}, X.~{Mu}, and L.~{Hanzo}, ``{Near-Field Communications: What Will Be Different?}'' \emph{arXiv e-prints}, p. arXiv:2303.04003, Mar. 2023.

\bibitem{opportunistic}
K.-K. Wong, K.-F. Tong, Y.~Chen, Y.~Zhang, and C.-B. Chae, ``Opportunistic fluid antenna multiple access,'' \emph{IEEE Transactions on Wireless Communications}, vol.~22, no.~11, pp. 7819--7833, 2023.

\bibitem{beam_focusing}
H.~Zhang, N.~Shlezinger, F.~Guidi, D.~Dardari, M.~F. Imani, and Y.~C. Eldar, ``Beam focusing for near-field multiuser {MIMO} communications,'' \emph{IEEE Transactions on Wireless Communications}, vol.~21, no.~9, pp. 7476--7490, 2022.

\bibitem{Multiple-access}
Z.~Wu and L.~Dai, ``Multiple access for near-field communications: Sdma or ldma?'' \emph{IEEE Journal on Selected Areas in Communications}, vol.~41, no.~6, pp. 1918--1935, 2023.

\bibitem{unified}
H.~Lu and Y.~Zeng, ``Communicating with extremely large-scale array/surface: Unified modeling and performance analysis,'' \emph{IEEE Transactions on Wireless Communications}, vol.~21, no.~6, pp. 4039--4053, 2022.

\bibitem{RIS-4096}
H.~Yan, H.~Chen, W.~Liu, S.~Yang, G.~Wang, and C.~Yuen, ``Ris-enabled joint near-field 3d localization and synchronization in siso multipath environments,'' \emph{IEEE Transactions on Green Communications and Networking}, pp. 1--1, 2024.

\bibitem{near_field_MU}
H.~Lu and Y.~Zeng, ``Near-field modeling and performance analysis for multi-user extremely large-scale {MIMO} communication,'' \emph{IEEE Communications Letters}, vol.~26, no.~2, pp. 277--281, 2022.

\bibitem{RIS-2873}
Y.~Pan, C.~Pan, S.~Jin, and J.~Wang, ``Ris-aided near-field localization and channel estimation for the terahertz system,'' \emph{IEEE Journal of Selected Topics in Signal Processing}, vol.~17, no.~4, pp. 878--892, 2023.

\bibitem{wideband-nearfield}
M.~Cui and L.~Dai, ``Near-field wideband beamforming for extremely large antenna arrays,'' 2023.

\bibitem{Parallel}
Z.~Wang, X.~Mu, Y.~Liu, and R.~Schober, ``Ttd configurations for near-field beamforming: Parallel, serial, or hybrid?'' \emph{IEEE Transactions on Communications}, vol.~72, no.~6, pp. 3783--3799, 2024.

\bibitem{rainbow}
M.~Cui, L.~Dai, Z.~Wang, S.~Zhou, and N.~Ge, ``Near-field rainbow: Wideband beam training for xl-{MIMO},'' \emph{IEEE Transactions on Wireless Communications}, vol.~22, no.~6, pp. 3899--3912, 2023.

\bibitem{DAS1}
X.-H. You, D.-M. Wang, B.~Sheng, X.-Q. Gao, X.-S. Zhao, and M.~Chen, ``Cooperative distributed antenna systems for mobile communications [coordinated and distributed {MIMO}],'' \emph{IEEE Wireless Communications}, vol.~17, no.~3, pp. 35--43, 2010.

\bibitem{DAS2}
S.~Zhou, M.~Zhao, X.~Xu, J.~Wang, and Y.~Yao, ``Distributed wireless communication system: a new architecture for future public wireless access,'' \emph{IEEE Communications Magazine}, vol.~41, no.~3, pp. 108--113, 2003.

\bibitem{DAS3}
S.~Ye, M.~Xiao, M.-W. Kwan, Z.~Ma, Y.~Huang, G.~Karagiannidis, and P.~Fan, ``Extremely large aperture array {(ELAA)} communications: Foundations, research advances and challenges,'' \emph{IEEE Open Journal of the Communications Society}, vol.~5, pp. 7075--7120, 2024.

\bibitem{LoS_MIMO}
H.~Do, S.~Cho, J.~Park, H.-J. Song, N.~Lee, and A.~Lozano, ``Terahertz line-of-sight {MIMO} communication: Theory and practical challenges,'' \emph{IEEE Communications Magazine}, vol.~59, no.~3, pp. 104--109, 2021.

\bibitem{LoS_MIMO2}
H.-J. Song and N.~Lee, ``Terahertz communications: Challenges in the next decade,'' \emph{IEEE Transactions on Terahertz Science and Technology}, vol.~12, no.~2, pp. 105--117, 2022.

\bibitem{sparsearray1}
M.~Lou, J.~Jin, H.~Wang, D.~Wu, L.~Xia, Q.~Wang, Y.~Yuan, and J.~Wang, ``Performance analysis of sparse array based massive {MIMO} via joint convex optimization,'' \emph{China Communications}, vol.~19, no.~3, pp. 88--100, 2022.

\bibitem{modular}
X.~Li, H.~Lu, Y.~Zeng, S.~Jin, and R.~Zhang, ``Modular extremely large-scale array communication: Near-field modelling and performance analysis,'' \emph{China Communications}, vol.~20, no.~4, pp. 132--152, 2023.

\bibitem{MU-modular}
X.~Li, Z.~Dong, Y.~Zeng, S.~Jin, and R.~Zhang, ``Multi-user modular {XL-{MIMO}} communications: Near-field beam focusing pattern and user grouping,'' \emph{IEEE Transactions on Wireless Communications}, vol.~23, no.~10, pp. 13\,766--13\,781, 2024.

\bibitem{HSPM}
Y.~Chen, L.~Yan, and C.~Han, ``Hybrid spherical- and planar-wave modeling and {DCNN}-powered estimation of terahertz ultra-massive {MIMO} channels,'' \emph{IEEE Transactions on Communications}, vol.~69, no.~10, pp. 7063--7076, 2021.

\bibitem{Theory}
H.~Do, S.~Cho, J.~Park, H.-J. Song, N.~Lee, and A.~Lozano, ``Terahertz line-of-sight {MIMO} communication: Theory and practical challenges,'' \emph{IEEE Communications Magazine}, vol.~59, no.~3, pp. 104--109, 2021.

\bibitem{ZF}
Q.~Spencer, A.~Swindlehurst, and M.~Haardt, ``Zero-forcing methods for downlink spatial multiplexing in multiuser {MIMO} channels,'' \emph{IEEE Transactions on Signal Processing}, vol.~52, no.~2, pp. 461--471, 2004.

\bibitem{HSPM_IRS}
Y.~Chen, R.~Li, C.~Han, S.~Sun, and M.~Tao, ``Hybrid spherical- and planar-wave channel modeling and estimation for terahertz integrated {UM-MIMO} and {IRS} systems,'' \emph{IEEE Transactions on Wireless Communications}, vol.~22, no.~12, pp. 9746--9761, 2023.

\bibitem{HDA}
F.~Sohrabi and W.~Yu, ``Hybrid digital and analog beamforming design for large-scale antenna arrays,'' \emph{IEEE Journal of Selected Topics in Signal Processing}, vol.~10, no.~3, pp. 501--513, 2016.

\end{thebibliography}
\end{document}